# Phase dynamics in a stack of inductively coupled intrinsic Josephson junctions and terahertz electromagnetic radiation

Xiao Hu and Shi-Zeng Lin

World Premier International Center for Materials Nanoarchitectonics, National Institute for Materials Science, Tsukuba 305-0044, Japan

Japan Science and Technology Agency, Kawaguchi 332-0012, Japan

**Abstract**

The Josephson effect is a phenomenon of current flow across two weakly linked superconductors separated by a thin barrier, i.e. Josephson junction, associated with coherent quantum tunneling of Cooper pairs. Many novel phenomena appear due to the nonlinear property of the Josephson effect, such as Shapiro steps in dc current-voltage (*IV*) characteristics of a Josephson junction under microwave shining, which can be used as a voltage standard. The Josephson effect also provides a unique way to generate high-frequency electromagnetic (EM) radiation by dc bias voltage as the inverse process of the Shapiro step. The discovery of cuprate high-$T_c$ superconductors accelerated the effort to develop novel source of EM waves based on a stack of atomically dense-packed intrinsic Josephson junctions (IJJs), since the large superconductivity gap covers the whole terahertz (THz) frequency band. Very recently, strong and coherent THz radiations have been successfully generated from a mesa structure of $Bi_2Sr_2CaCu_2O_{8+\delta}$ single crystal which works both as the source of energy gain and as the cavity for resonance. This experimental breakthrough posed a challenge to theoretical study on the phase dynamics of stacked IJJs, since the phenomenon cannot be explained by the known solutions of the sine-Gordon equation so far. It is then found theoretically that, due to huge inductive coupling of IJJs produced by the nanometer junction separation and the large London penetration depth of order of $\mu$m of the material, a novel dynamic state is stabilized in the coupled sine-Gordon system, in which $\pm\pi$ kinks in phase differences are developed responding to the standing wave of Josephson plasma and are stacked alternately in the *c*-axis. This novel solution of the inductively coupled sine-Gordon equations captures the important features of experimental observations. The theory predicts an optimal radiation power larger than the one observed in recent experiments by orders of magnitude, and thus suggests the technological relevance of the phenomena.

**I. Introduction**

In 1962 Josephson predicted that the current flow associated with coherent quantum tunneling of Cooper pairs through a barrier between two superconducting electrodes, i.e. Josephson junction, is proportional to the sinusoidal function of the phase difference, known as the dc Josephson relation, and that the time derivative of phase difference is proportional to the voltage across the barrier, known as the ac Josephson relation [1-3]. The Josephson effect causes many unique, highly nonlinear phenomena. For example, when a Josephson junction is irradiated by a microwave, plateaus of constant voltage called Shapiro steps appear in the current-volgate (*IV*) characteristics [4, 5] at a sequence of equally separated values of dc bias voltage. Since the separation of voltages is given by $\omega\hbar/2e$ with $\omega$ the angular frequency of the microwave, $\hbar$ the Planck constant and $e$ the charge of electron, the Shapiro steps are used as a voltage standard [6]. Just in the opposite direction, the Josephson effect provides a unique principle to excite high-frequency electromagnetic (EM) wave [1, 2, 7], which was first confirmed in single junctions, and then in Josephson junction arrays.

Since the discovery of the cuprate high-$T_c$ superconductors of layered structures[8], and the demonstation of the intrinsic Josephson effect in $Bi_2Sr_2CaCu_2O_{8+\delta}$ (BSCCO) [9], the effort to explore a strong EM source based on the nano-scale, built-in Josephson junctions has been accelerated [10, 11]. These so-called intrinsic Josephson junctions (IJJs) have obvious advantages superior to the artificial ones since the former are homogeneous at the atomic scale guaranteed by the high quality of single crystals, and the superconductivity gap is large, typically of tens of meV, which permits the frequency to cover the whole range of the terahertz (THz) band, a very useful regime of EM waves still lacking of compact solid-state generators [12, 13]. However, it was difficult to synchronize the whole stack of IJJs which is crucial for increasing the radiation power.

An experimental breakthrough was achieved in 2007 [14]. Clear evidences have been obtained that coherent THz radiations from a rectangular mesa mounted on the top of a substrate of a single crystal of BSCCO were realized by a *c*-axis bias voltage in the absence of external magnetic field. The strong EM emission takes place at the bias voltage when the frequency determined by the ac Josephson relation equals to the fundamental cavity mode, corresponding to a half wavelength of the Josephson plasma in the mesa [14]. The experimental observations cannot be understood with the conventional wisdom on Josephson phenomena developed mainly based on single Josephson junctions, since the known cavity modes associated with finite external magnetic field are clearly irrelevant and solitons, which can be excited in the absence of external magnetic field and correspond to cavity modes, cannot be stacked uniformly in $c$-axis to achieve in-phase dynamics in all the junctions. It is also not clear why it is possible to synchronize the superconductivity phase differences of totally ~600 junctions.

Motivated by this recent experiment, theoretical investigations on the sine-Gordon equations for the IJJs have been carried out with special focus on the case of strong inductive coupling originated in the extremely small junction spacing of $s=1.5\text{nm}$ as compared with the London penetration depth $\lambda_{ab} \approx 400\text{nm}$. A new dynamic state of superconductivity phase difference has been found for the system under a dc voltage bias. The system creates $\pm\pi$ phase kinks when the phase differences rotate overall with the angular velocity determined by the bias voltage according to the ac Josephson relation. The $\pi$ phase kinks couple the lateral cavity modes of the transverse Josephson plasma to the dc bias voltage in an efficient way. When the bias voltage is tuned to the value corresponding to the cavity frequency of the Josephson plasma, a cavity resonance takes place and the amplitude of plasma is enhanced significantly. A large dc Josephson current appears and injects a large amount of energy into the system under the dc voltage bias, and a part of the energy is radiated as EM wave from the sides of mesa into space. The alternating arrangement of $\pm\pi$ phase kinks in the IJJs generates a strong attractive interaction between neighboring junctions, which compensates the self-energy of the phase kinks and thus keeps the total energy of the state low.

The state mentioned above provides a possible candidate responsible for the experimental observations on strong and coherent THz radiations based on IJJs of cuprates. Needless to say the final knowledge has not been achieved yet. The aim of this review is to assemble the novel physical phenomena caused by the Josephson effects in both single and stack of Josephson junctions, and to compose a consistent picture for them based on the present understanding, which is expected to pave a better access to the field.

This review article is organized as follows. In Sec II, we begin with the principle of Josephson effects, and discuss the Josephson phenomena in single Josephson junctions, with which the basic concepts can be best illustrated. Experimental efforts for realization of EM radiation based on the artificial Josephson junctions are briefly summarized thereafter. We then proceed to Sec. III to introduce recent experimental progresses in stimulating THz EM radiation based on IJJ stacks in the absence of external magnetic field. The coupled sine-Gordon equations for the phase dynamics of IJJs are then analyzed with the focus on the case of strong inductive coupling. The $\pi$ kink state is then derived and is shown to be able to explain the main features of the experimental observation. Stimulated EM radiations based on IJJs with an external magnetic field are discussed in Sec. IV, and several superconductivity-based techniques to excite EM waves other than the Josephson effect are reported in Sec. V. Conclusion is given in Sec. VI with some perspectives.

**II Josephson effects and single Josephson junctions**

*II.1 Josephson effects*

A macroscopic system in superconducting state can be described by a single wave function [15, 16]:

$$\Psi = |\Psi| e^{i\varphi} \quad . \quad (2.1)$$

When two superconductors get close to each other, Cooper pairs can tunnel through the barrier in between, which makes the two subsystems intervene with each other. In terms of the gauge invariant phase difference

$$\gamma = \varphi_a - \varphi_b - \frac{2\pi}{\phi_0} \int_b^a \mathbf{A} \cdot d\mathbf{s} \quad , \quad (2.2)$$

where $\phi_0 = hc/2e$ is the flux quantum, with $h$ the Plank constant, $c$ the light velocity in vacuum and $e$ the electron charge, the current flow between the two superconductors is given by the dc Josephson relation [1, 2]

$$J = J_c \sin \gamma , \quad (2.3)$$

with $J_c$ the critical current density. When a voltage is applied across the barrier, the phase difference evolves with time following the ac Josephson relation [1, 2]

$$2eV = \hbar \frac{d\gamma}{dt} . \quad (2.4)$$

The critical current $J_c$ depends on the applied voltage (hence the plasma frequency) according to a microscopic theory developed by Ambegaokar and Baratoff [17]. The dependence is very weak for $2eV \ll \Delta$ with $\Delta$ the superconductivity gap, and thus is neglected in the discussion hereafter. The total system including two superconducting electrodes and the barrier where the above Josephson phenomena take place is generally called Josephson junction.

*II.2 sine-Gordon equation for Josephson junctions*

The above Josephson relations impose rules for the spatial and temporal variations of the gauge invariant phase difference, since the electric and magnetic fields themselves follow the Maxwell equations. The spatial variation of gauge invariant phase difference across a Josephson junction with separation $d$ along at the $z$ direction is given by

$$\nabla \gamma = \frac{2\pi(2\lambda+d)}{\phi_0} \mathbf{B} \times \hat{\mathbf{z}} \quad , \qquad (2.5)$$

where $\nabla$ is the gradient operator in lateral directions, $\lambda$ is the London penetration depth for an isotropic superconductor, $\hat{\mathbf{z}}$ the unit vector along the $z$ direction. With the Maxwell equation

$$\nabla \times \mathbf{B} = \frac{4\pi}{c}\mathbf{J} + \frac{\varepsilon}{c}\frac{\partial \mathbf{E}}{\partial t} \quad , \qquad (2.6)$$

where $\mathbf{B}$ is the magnetic field, $\mathbf{J}$ the current, $\mathbf{E}$ the electric field and $\varepsilon$ the relative dielectric constant of the material of barrier. Using the current relation

$$J^z = J_c \sin\gamma + \sigma V/d \quad , \qquad (2.7)$$

where $\sigma$ is the conductivity carried by quasiparticles, we arrive at the following sine-Gordon equation [7]:

$$\frac{\partial^2 \gamma}{\partial x^2} + \frac{\partial^2 \gamma}{\partial y^2} - \frac{1}{c'^2}\frac{\partial^2 \gamma}{\partial t^2} - \frac{\beta'}{c'^2}\frac{\partial \gamma}{\partial t} = \frac{1}{\lambda_J^2}\sin\gamma \quad . \qquad (2.8)$$

Here

$$c' = c/\sqrt{\varepsilon(1+2\lambda/d)} \qquad (2.9)$$

is the Swihart velocity [18],

$$\lambda_J = \sqrt{c\phi_0 / \left[8\pi^2 J_c(d+2\lambda)\right]} \qquad (2.10)$$

the Josephson penetration depth, and

$$\beta' = 4\pi\sigma/\varepsilon(=1/RC), \qquad (2.11)$$

with $R$ the resistivity and $C$ the capacitance of the junction. The Swihart velocity is much smaller than the velocity of light in the material of barrier due to the deformation of the EM wave, since the electric field is confined in the small separation of junction while the magnetic field can penetrate into the superconductor over the scale of penetration depth. Typically, $\lambda_J \sim 1\text{mm}$ and $c' \sim c/20$ for Josephson junctions made of conventional low-temperature superconductors.

We first concentrate on the static case, and for simplicity consider the case of one dimension, where the sine-Gordon equation (2.8) is reduced to

$$\frac{d^2\gamma}{dx^2} = \frac{1}{\lambda_J^2}\sin\gamma \quad . \qquad (2.12)$$

Due to the induced magnetic field, the Josephson current is confined in a scale of $\lambda_J$ from the edge of junction. For a short junction with the lateral size smaller than $\lambda_J$ where the screening effect can be neglected, $\gamma$ shows a linear spatial variation according to a static external magnetic field, which then modulates the total Josephson current in a way showing the Fraunhofer pattern [19].

For a long junction, the screening effect should be treated in order to understand the true behavior of the system. When the external magnetic field is small, it is confined at the edges due to the screening effect of the Josephson current. When the external magnetic field is large enough, it starts to penetrate into the whole junction, which causes a $2\pi$ variation in the phase difference, similar to the Abrikosov vortex in a type II superconductor. The analogy between the Josephson junction and the pendulum is useful [20], where the kinetic energy of pendulum corresponds to the spatial derivative of the gauge invariant phase difference and the potential energy of pendulum corresponds to the Josephson energy. The critical magnetic field is then evaluated as the minimal kinetic energy necessary for a complete rotation of the pendulum

$$H_{c1} = \frac{\phi_0}{\pi \lambda_J (2\lambda + d)} \quad . \quad (2.13)$$

A Josephson vortex does not carry a core as an Abrikosov vortex, where the amplitude of superconductivity order parameter is suppressed to zero.

*II.3 Josephson plasma*

At small value of $\gamma$, the system (2.8) can show a back-and-forth oscillation of Cooper pairs between the two superconductor electrodes with the frequency

$$\omega_J = c'/\lambda_J, \quad (2.14)$$

which is accompanied by a uniform z-axis electric field and zero magnetic field. $\omega_J$ is called Josephson plasma frequency, and it has typical order of 10GHz for junctions made of low-temperature superconductors. This longitudinal plasma is the Nambu-Goldstone mode associated with the spontaneous (local) gauge symmetry breaking of the superconductivity transition. The mode is suppressed when temperature approaches the transition point from below since the diverging penetration depth $\lambda$ reduces the frequency to zero.

There exists also a transverse plasma waves carrying lateral variation of phase difference, which exhibits the following dispersion relation (see Figure 1)

$$\omega^2 = \omega_J^2 + k^2 c'^2 \quad . \quad (2.15)$$

The frequency of travelling plasma wave should be above $\omega_J$, which indicates that the Josephson junction is transmittable only for EM waves of high frequency. As can be seen from the sine-Gordon equation (2.8), only when the frequency is beyond $\omega_J$ the effective screening length becomes imaginary, which permits the Josephson plasma to propagate.

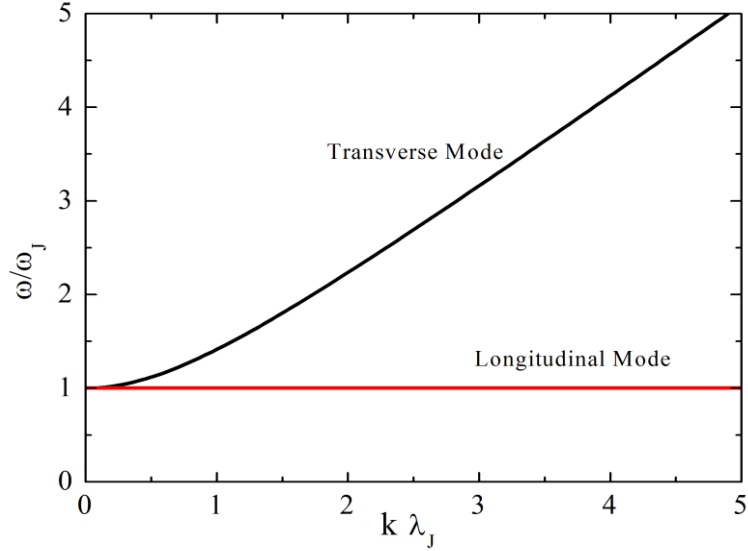

Figure 1. Plasma modes in a single Josephson junction.

*II.4 Current-Voltage (IV) characteristics*

Consider a short junction where the screening effect can be neglected. The gauge invariant phase difference can be taken as a constant except for a term associated with the constant bias current $J_{ext}$,

$$\gamma_0 = \frac{x^2 + y^2}{4\lambda_J^2} \frac{J_{ext}}{J_c} \quad . \quad (2.16)$$

The sine-Gordon equation (2.8) then is reduced to

$$\frac{J_{ext}}{J_c} = \sin\gamma + \frac{1}{\omega_J^2}\frac{\partial^2 \gamma}{\partial t^2} + \frac{\beta'}{\omega_J}\frac{1}{\omega_J}\frac{\partial \gamma}{\partial t} \quad , \quad (2.17)$$

or equivalently

$$J_{ext} = J_c \sin\gamma + \frac{\hbar}{2e}C\frac{d^2\gamma}{dt^2} + \frac{\hbar}{2e}\frac{\sigma}{d}\frac{d\gamma}{dt} \quad . \quad (2.18)$$

This can be directly read as a parallel circuit consisted from a nonlinear Josephson junction, a capacitor and a resistor. This RC shunted junction (RCSJ) model is useful for the understanding of *IV* characteristics of short Josephson junctions.

Equation (2.18) is the same as the equation of motion of a classical particle in a washboard potential $\cos\gamma$ tilted by the external current, with the capacitance and conductivity corresponding to the mass and viscous drag. Increasing the external current from zero, the system exhibits zero voltage as far as the current is smaller than the critical current. For negligibly small capacitance, the system is governed by the viscous drag force and the voltage increases continuously with current beyond the critical current, and asymptotically approaches the linear *IV* line given by the conductivity. For sufficiently large capacitance, the system behaves hysterically due to the inertia effect: on increasing current the voltage jumps discontinuously from zero to finite value at the critical current; on decreasing current the voltage remains finite even below the critical current, and becomes zero only at a re-trapping current. The re-trapping current approaches zero for large capacitance. The *IV* curves for several typical values of $\beta'$ are shown in Figure 2.

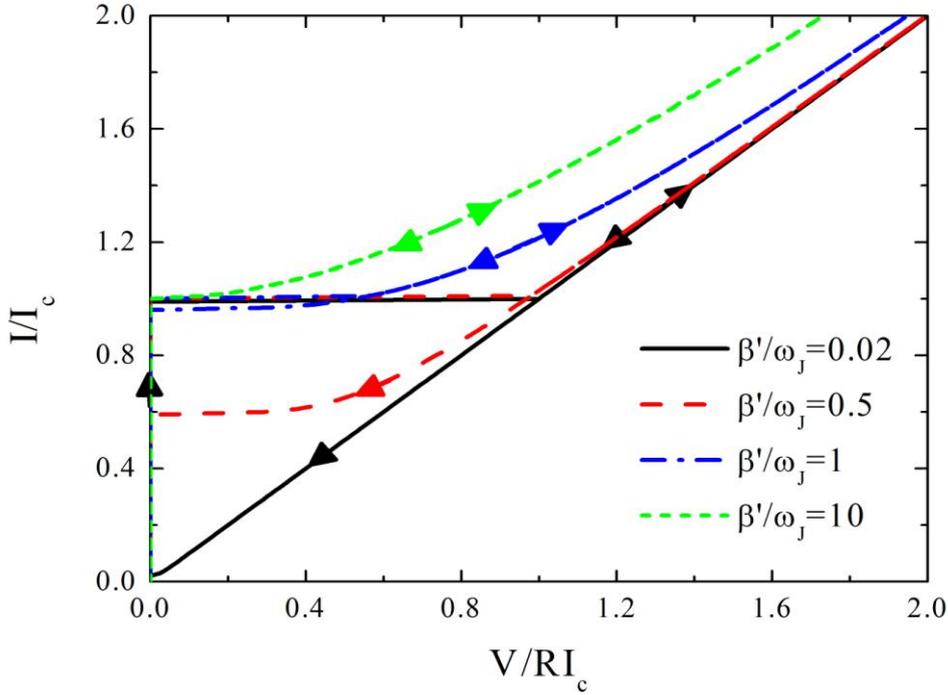

Figure 2. Current-voltage curves for several typical values of β' obtained by numerical calculations. Arrows indicate the direction of current sweeping.

*II.5 Resonant mode: Fiske step*

Now we consider a general case where a parallel magnetic field and a dc voltage are applied to the Josephson junction. We suppose that the magnetic field induced by the current is negligible compared with the external one. For a junction of finite length, resonances take place when the frequency determined by the voltage according to the ac Josephson relation matches one of the standing modes of the cavity.

For an open-end cavity (magnetic field is zero at the boundary), the voltage takes the following form for the $n$-th mode of standing wave

$$v_n = \cos\frac{n\pi x}{L} \quad , \quad (2.19)$$

with a frequency $\omega_n = n\pi c'/L$, where $L$ is the length of the junction. In the following analysis, we assume the time dependence of the electromagnetic fields and phase difference to be essentially determined by the dc voltage as

$$\omega = 2eV_0/\hbar . \quad (2.20)$$

The phase difference takes the form

$$\gamma(x,t) = \omega t - kx + \gamma_1(x,t) \quad , (2.21)$$

where

$$k = \frac{2\pi(2\lambda + d)}{\phi_0} H \quad (2.22)$$

is a quasi wave number introduced by the external magnetic field. When the term $\gamma_1(x,t)$ is small, one arrives at the following equation from Eq.(2.8)

$$\frac{\partial^2 \gamma_1}{\partial x^2} - \frac{1}{c'^2}\frac{\partial^2 \gamma_1}{\partial t^2} - \frac{\beta'}{c'^2}\frac{\partial \gamma_1}{\partial t} = \frac{1}{\lambda_J^2}\sin(\omega t - kx) \quad , (2.23)$$

where we suppose the solution is uniform along the direction of the external magnetic field (here the $y$ direction). From the ac Josephson relation, one has the following expansion of $\gamma_1(x,t)$ in terms of cavity modes

$$\gamma_1(x,t) = \text{Im}\left\{ e^{i\omega t}\sum_{n=0}^{\infty} g_n \cos\frac{n\pi x}{L}\right\} \quad , \quad (2.24)$$

where $g_n$ are complex numbers. Inserting Eq.(2.24) into Eq.(2.23), one arrives at

$$\sum_{n=0}^{\infty} g_n \cos\frac{n\pi x}{L}\left[\frac{\omega^2}{c'^2} - \left(\frac{n\pi}{L}\right)^2 - i\frac{\omega\beta'}{c'^2}\right] = \frac{1}{\lambda_J^2} e^{-ikx} \quad . (2.25)$$

Multiplying by $\cos(n\pi x/L)$ and integrating, we obtain

$$g_n = \frac{2c'^2}{\omega^2 \lambda_J^2}\frac{1-(\omega_n/\omega)^2 + i/Q}{\left[1-(\omega_n/\omega)^2\right]^2 + 1/Q^2}[B_n - iC_n] \quad , \quad (2.26)$$

with the quality factor $Q = \beta'/\omega$ and

$$B_n = \frac{1}{L}\int_0^L dx \cos\frac{n\pi x}{L}\cos kx \quad , \quad (2.27)$$

$$C_n = \frac{1}{L}\int_0^L dx \cos\frac{n\pi x}{L}\sin kx \quad . \quad (2.28)$$

The dc current carried by the Josephson junction then is given by

$$J_{dc} = \lim_{T \to \infty} \frac{1}{T} \int_0^T dt \frac{1}{L} \int_0^L dx J_c \sin(\omega t - kx + \gamma_1)$$

$$= \frac{c'^2 J_c}{\omega^2 \lambda_J^2} \sum_{n=0}^{\infty} \frac{1/Q}{\left[1-(\omega_n/\omega)^2\right]^2 + 1/Q^2} \left[\frac{kL}{kL+n\pi} \times \frac{\sin(kL/2-n\pi/2)}{kL/2-n\pi/2}\right]^2 \quad . \quad (2.29)$$

When the voltage matches one of the cavity frequencies of the junction, the dc current is enhanced. These singularities in the *IV* characteristic were first observed by Fiske (1964) [21], and now known as the Fiske steps. For a given external magnetic field, the Fiske step is most pronounced when

$$\omega = kc', \quad (2.30)$$

namely at the voltage

$$V_0 = \frac{2\lambda + d}{\sqrt{\varepsilon}} H \quad . \quad (2.31)$$

The energy input by the dc voltage is dissipated into Joule heating, partially in a nonlinear way caused by the Josephson oscillation as described by Eq.(2.29), and partially in the trivial ohmic way which is not included here. It is clear from Eq.(2.29) that cavity resonances occur only when an external magnetic field is present. For a zero external magnetic field, only the uniform mode $n=0$ exists and one has

$$J_{ext} = \frac{1}{2} \frac{c'^2 J_c}{\lambda_J^2} \frac{\beta'}{\omega^3 + \beta'^2 \omega} \quad . (2.32)$$

*II.6 Resonant mode: Eck peak*

For a junction of effectively infinite length, there is no standing wave. A resonance occurs when the wave number of Josephson plasma matches the quasi wave length induced by the external magnetic field. In this case, instead of Eq.(2.24), one should have

$$\gamma_1(x,t) = \text{Im}\left\{g(k,\omega)e^{i(\omega t - kx)}\right\}, \quad (2.33)$$

where

$$g(k,\omega) = \frac{1/\lambda_J^2}{\omega^2/c'^2 - k^2 - i\beta'\omega/c'^2} \quad . \quad (2.34)$$

The dc current is then given by

$$J_{dc} = \frac{J_c c'^2}{2\lambda_J^2 \omega^2} \frac{1/Q}{\left(1 - k^2 c'^2/\omega^2\right)^2 + 1/Q^2} \quad . \quad (2.35)$$

There is only one enhancement of dc current in the *IV* characteristics at the position given by Eq.(2.31), which is first observed by Eck *et al* (1964) [22] and now known as the Eck peak.

*II.7 Soliton solution*

Let us consider the sine-Gordon equation (2.8) in one dimension without dissipation and power input:

$$\frac{\partial^2 \gamma}{\partial x^2} - \frac{\partial^2 \gamma}{\partial t^2} = \sin\gamma \quad , (2.36)$$

where the coordinate $x$ and time $t$ are measured in units of $\lambda_J$ and $\lambda_J/c'$. Since the above equation cannot be solved generally, we concentrate here on the case in which a wave is propagating with the unchanged form, for example towards the positive direction of $x$ axis [23-25]

$$\gamma = \gamma(x - ut) \quad (2.37)$$

with $u$ an arbitrary, constant propagation velocity. The equation then is reduced to

$$\frac{d^2\gamma}{d\xi^2} = \frac{\sin\gamma}{1-u^2} \quad . (2.38)$$

One can integrate it as

$$\frac{d\gamma}{d\xi} = \sqrt{\frac{2(E-\cos\gamma)}{1-u^2}} \quad (2.39)$$

with $E$ an integral constant. There are many different solutions to the above equation, depending on $E$ and the propagation velocity. For $E=1$ (one must have $|u|<1$), the soliton solution is realized

$$\gamma = 4\tan^{-1}\left[\exp\left(\frac{x-ut}{\sqrt{1-u^2}}\right)\right] \quad , (2.40)$$

corresponding to a rotation of $\gamma$ from 0 by $2\pi$. The soliton and associated magnetic field correspond to Eq. (2.40) are shown in Figure 3.

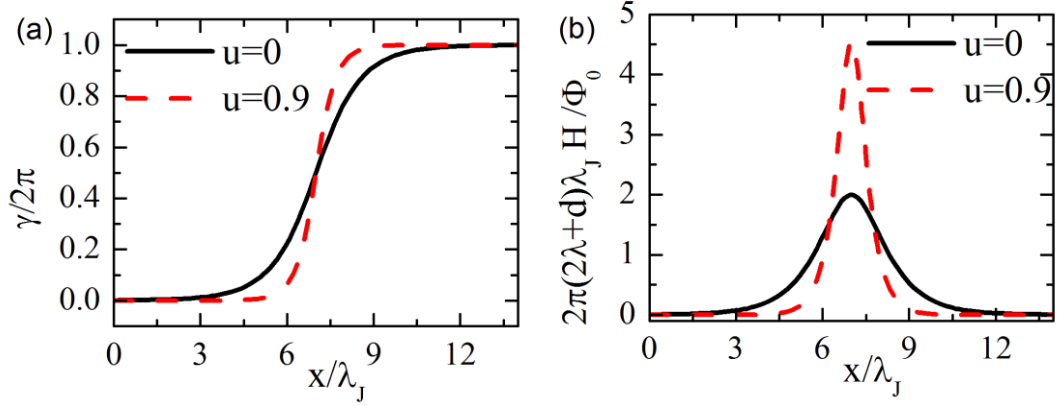

Figure 3. (a) Solitons at different velocities, and (b) magnetic fields associated with the solitons.

*II.8 Zero-field Fiske steps*

In an ideal junction a soliton described by Eq. (2.36) can travel at any velocity below the plasma velocity. In the presence of dissipation, the soliton loses energy when travelling and the associated phase kink smears out. The dissipation can be compensated by an input power and the soliton moves at a velocity determined by the power balance condition, i.e. the dissipation is compensated by the dc input power [26]. In a junction with finite length, the soliton interacts with the environment in a complicated manner [27]. For the open boundary condition with zero magnetic field, the soliton is reflected at the junction edge into an anti-soliton. For the fixed boundary condition where the superconductivity phase is pinned by defects or microshorts, the reflected soliton has the same polarity as the incident one [28].

For the open boundary condition, the voltage pulse generated by the motion of soliton and reflected anti-soliton has the same polarity, thus a non-zero average voltage is induced. Although there is no exact solution to the sine-Gordon equation with dissipation, as a good approximation one can evaluate the average voltage, input dc current, and the period of the soliton motion as the function of the soliton velocity $v$ and the junction length [29],

$$\langle V \rangle = \frac{\phi_0}{L} \frac{v}{c} \quad , (2.41)$$

$$\langle V \rangle I = \int_0^L \sigma V^2 dx = \sigma \left(\frac{\hbar v}{2e}\right)^2 \frac{8}{\lambda_J \sqrt{1-(v/c')^2}} \quad , (2.42)$$

where it is assumed that the energy input is dissipated totally in the ohmic way. It is clear from Eq.(2.42) that when the velocity of soliton approaches the light velocity in the material, and consequently the voltage approaches

$$\langle V \rangle = c'\phi_0/cL, \quad (2.43)$$

the junction can bypass very large current. This enhancement of current is known as the zero-field Fiske step, observed first by Chen *et al* (1971) [30]. A typical *IV* characteristics with zero-field Fiske steps is shown in Figure 4.

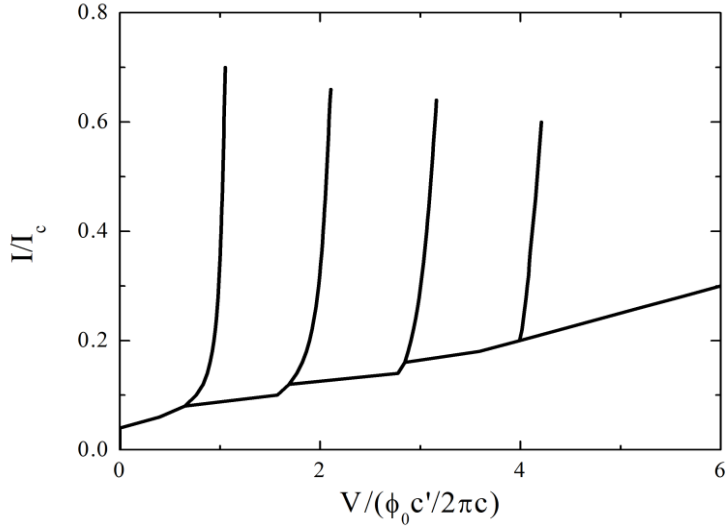

Figure 4. Zero-field Fiske steps for a single junction with L=6 caused by the periodic motion and reflection of solitons.

When the soliton hits the edge, an EM pulse is emitted into the space. The frequency of EM wave *f* is determined by the period of the shuttle motion of soliton, and is related to the average voltage according to Eq. (2.43) as

$$\langle V \rangle = 2f \frac{\phi_0}{c}. \quad (2.44)$$

It is worthy to notice that this frequency is half of that determined by the ac Josephson relation for the same given voltage

$$V = f \frac{\phi_0}{c}. \quad (2.45)$$

Multi-soliton solutions generate a sequence of zero-field Fiske steps at evenly spaced voltages (see Figure 4). The maximal number of solitons accommodated in the junction is roughly $L/\lambda_J$. During the process of reflection of soliton at the edges, a part of energy carried by the soliton is emitted in the form of EM wave. This radiation is avoided in ring-shaped Josephson junctions [31-33], which are expected to be useful for various applications such as a possible realization of quantum bit.

*II.9 Shapiro step*

When a dc voltage $V$ and an oscillating one $V_1 \cos \omega t$ are applied simultaneously across a junction, the phase difference of the junction evolves with time as

$$\gamma = \frac{2eV}{\hbar} t + \frac{2eV_1}{\omega \hbar} \sin \omega t + \gamma_0 \ , \quad (2.46)$$

where $\gamma_0$ is an arbitrary constant, and the junction is taken to be short and the phase difference is uniform. The Josephson current is given by

$$J = J_c \sin \gamma = J_c \sum_{n=-\infty}^{\infty} J_n\left(\frac{2eV_1}{\omega \hbar}\right) \sin\left[\left(\frac{2eV}{\hbar} - n\omega\right)t + \gamma_0\right], \quad (2.47)$$

where the Fourier-Bessel expansion is used

$$\sin(\varphi + A \sin \phi) = \sum_{n=-\infty}^{\infty} J_n(A) \sin(\varphi - n\phi) \quad (2.48)$$

with $J_n(x)$ the Bessel function of first kind of integer order. Therefore, a dc component of current appears at the dc voltage

$$V = n \frac{\omega \hbar}{2e}. \quad (2.49)$$

It is seen in the *IV* characteristics as a voltage plateau, known as Shapiro step now [4, 5]. A typical *IV* curve is shown in Figure 5. The height of the step is

$$\delta J = 2 J_c J_n\left(\frac{2eV_1}{\omega \hbar}\right). \quad (2.50)$$

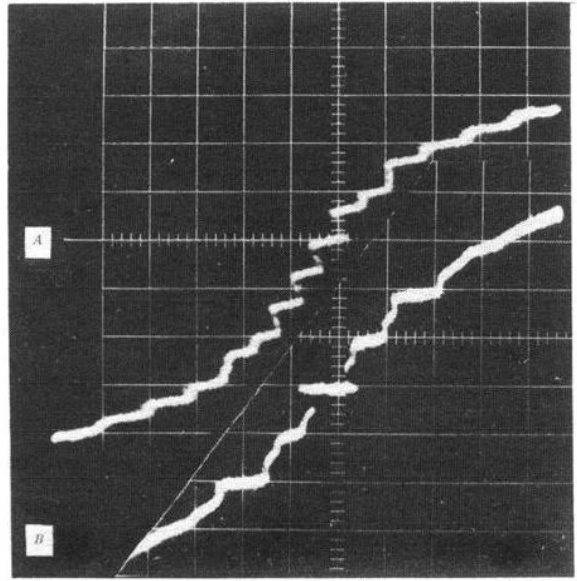

Figure 5. Shapiro steps in *IV* curve. The horizontal axis corresponds to current and the vertical

axis is voltage. [After S. Shapiro, Physical Review Letters 11, 80 (1963).]

*II.10 Electromagnetic Radiation based on the Josephson effects*

Observations on EM radiation from a Josephson junction date back to 1960s, shortly after the proposal of Josephson effects. The radiation at Fiske steps was detected by Yanson *et al* (1965) [34] and Langenberg *et al* (1965) [35]. The radiation caused by motion of solitons was reported by Dueholm *et al* (1981) [36]. The radiation from point contact [37, 38] and micro-bridge structures [39] which were placed in an external resonant cavity has also been reported. The radiation power in these experiments is of order of 1pW and the frequency is about 10GHz. After 40 years development, the cutting-edge technology of flux flow oscillators made of conventional single Josephson junctions has achieved radiation power about 10μW at frequency 600GHz, which has already been commercialized [40]. A typical radiation power of flux flow oscillators is shown in Figure 6.

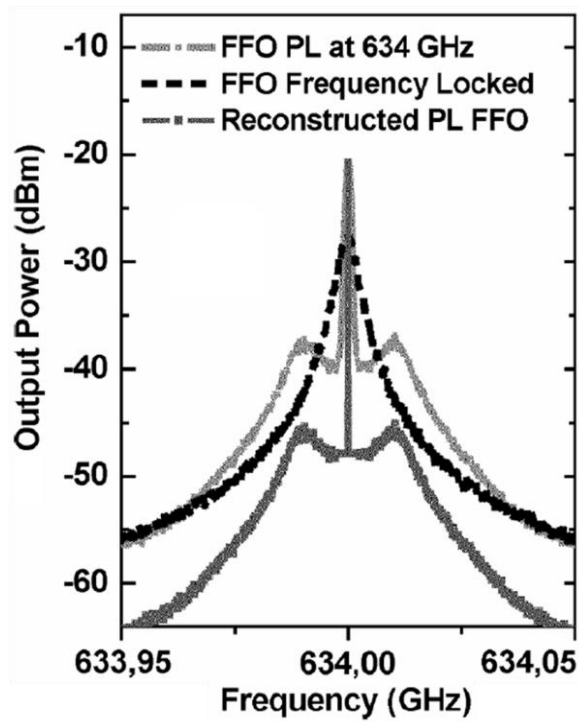

Figure 6. Spectra of the phase-locked (PL) and frequency locked flux flow oscillators (FFO); reconstructed spectra of the PL FFO are also shown. [After V. P. Koshelets *et al*, IEEE Transactions on Applied Superconductivity 15, 960-963 (2005).]

It is natural to accumulate Josephson junctions into array in order to derive strong radiation provided synchronization can be achieved. Josephson junctions may be either aligned on a plane and connected to each other to form two dimensional array [41-46] or piled up to form a stack [47-49]. One can also combine the two ways together to build an array of stacked Josephson junctions [50].

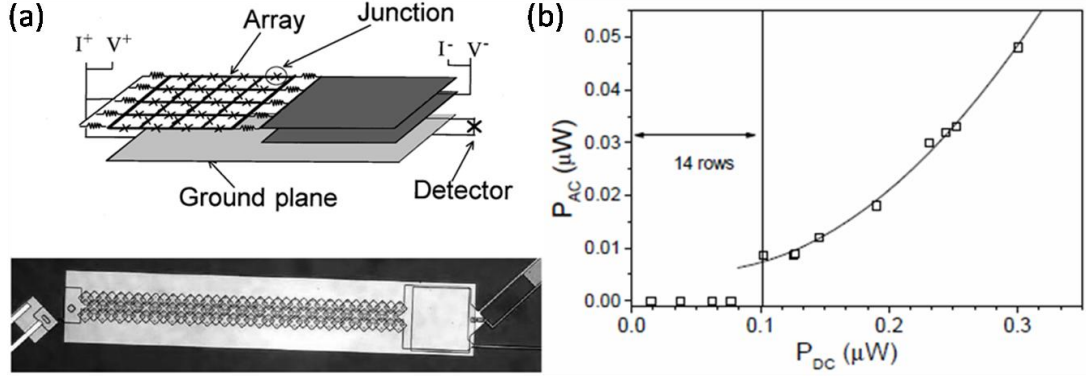

Figure 7. (a) Sketch of a typical sample (top). The square array, to the left, is capacitively coupled to a detector circuit, on the right. A ground plane underlies the entire structure. Picture of a 3 × 36 array (bottom). The length of the array is about 470 μm. (b) Detected ac power $P_{AC}$ as a function of the dc input power $P_{DC}$ for a 3 × 36 array (open squares). If all the non-zero voltage junctions are on the same dynamical state (resonant step, in this case), the input dc power is proportional to the number of these junctions. No output power is measured below 14 active rows. The solid line is the best quadratic fit for the data above 14 rows. For the units used in the plot, $P_{ac} = 0.007 - 0.07 P_{dc} + 0.7 P_{dc}^2$. [After P. Barbara et al, Physical Review Letters 82, 1963-1966 (1999).]

To make all the junctions operate in a synchronized way is one of the main challenges. One idea is to couple all Josephson junctions to a common resonator. This was demonstrated excellently by the experiment by P. Barbara et al (1999) [45]. A large number of Josephson junctions are integrated on a chip, and they are coupled to a passive high-Q resonator formed by the array itself and the ground plane, see Figure 7(a). When the population of active junctions reaches a threshold, strong emission appears with the radiation power increasing with the number of active junctions squared, in analogy to conventional lasers. The strong radiation induces current steps in the *IV* characteristics. As shown in Figure 7(b), the radiation frequency is 150GHz and the power is about 0.1μW, with the conversion efficiency from dc to ac power as high as 17%.

The mechanism for synchronization of the Josephson junction array has been studied based on the generalized Kuramoto model [51-54]. When the number of active junctions is small, the power in the common resonator fed by these active junctions is too small to force the junctions operate at the cavity frequency. Thus all junctions are running at their bare frequencies and own phases. However, as the population of active junctions increases, the cavity becomes powerful enough to synchronize all the junctions running coherently.

**III. Intrinsic Josephson Junctions of Cuprate High-$T_c$ Superconductors**

*III.1 Josephson effects in cuprate High-$T_c$ Superconductors*

The cuprate high-$T_c$ superconductors discovered in 80s of the last century possess commonly a layered structure [8]. Superconductivity takes place mainly on CuO layers, while the other components seem to play the role of providing carriers. So far, there is no well established theory for the mechanism for superconductivity in these strongly correlated electron systems. Nevertheless, as extremely type II superconductors with the London penetration depth much larger than the coherence length their vortex states [55] have been investigated intensively [56-60]

Due to the profound layer structure of the crystal, the cuprate superconductors can be modeled as a stack of Josephson junctions when the EM properties are concerned. This notion was demonstrated beautifully by Kleiner and Müller in BSCCO [9, 10]. The conductance of quasiparticle tunneling in BSCCO is very small and thus the junctions are in the underdamped regime. Each of the junctions can be either in the superconducting state or in the resistive state depending on the bias, which results in many branches in the *IV* characteristics, as shown in Figure 8.

The properties associated with the Josephson effects discussed in Sec. II for a single junction have also been observed in IJJ stacks, such as Fraunhofer pattern in critical current dependence on an applied magnetic field [9], Fiske resonances [61-63], and Shapiro steps [64-67]. For review articles on the intrinsic Josephson effects, please refer to [68, 69].
.

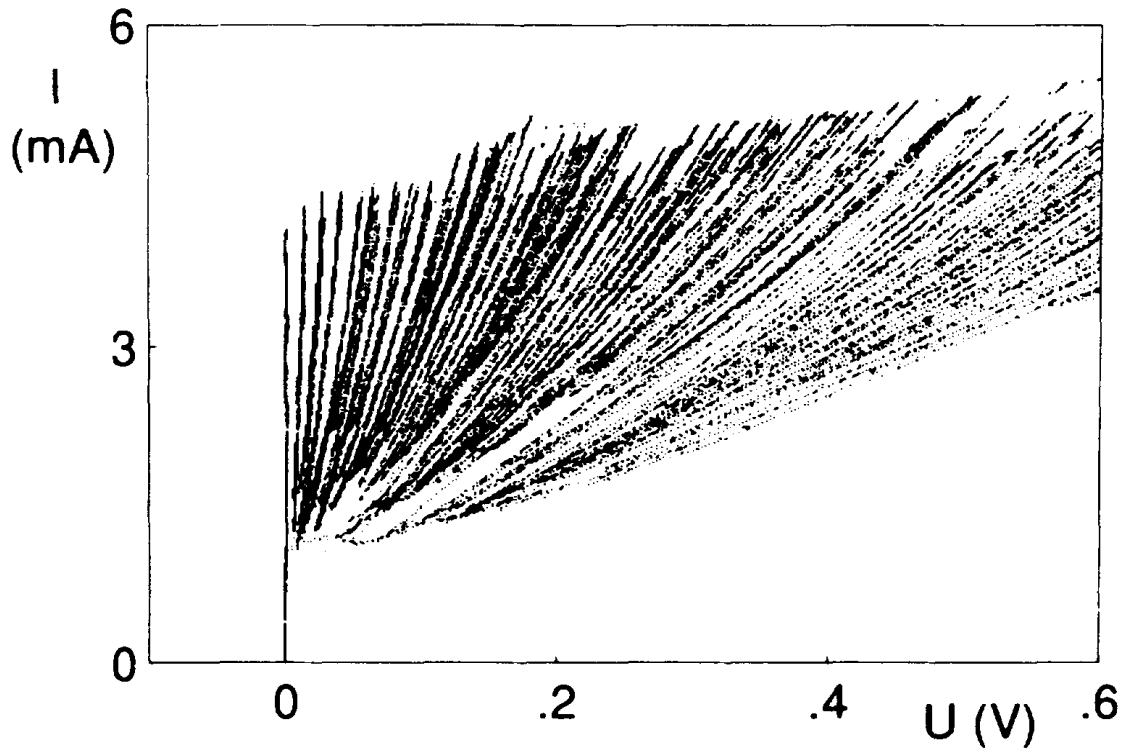

Figure 8. *IV* characteristics of BSCCO at 4.2K. The contact resistance of 18Ω is subtracted. [After R. Kleiner *et al*, Physical Review Letters 68, 2394-2397 (1992).]

*III.2 Inductively coupled sine-Gordon equations*

The Josephson junctions in cuprate high-$T_c$ superconductors are formed at atomic scale by the lattice structure, as such, the term of intrinsic Josephson junction was coined in order to make contrast to the ones made artificially. Here, two neighboring CuO layers and the block layer in between form one Josephson junction. Since the period of the junction stack is about 1.5nm, which is much smaller than the London penetration depth in the $c$-axis $\lambda_{ab} = 400$nm typically for underdoped BSCCO, superconductivity and EM dynamics cannot be screened locally. Therefore, the junctions are coupled to each other strongly due to the inductive interaction [70-72]. There exist other inter-junction couplings due to the charging effect [73, 74], and non-equilibrium effect [75, 76]. These couplings produce many peculiar properties in IJJs as discussed below.

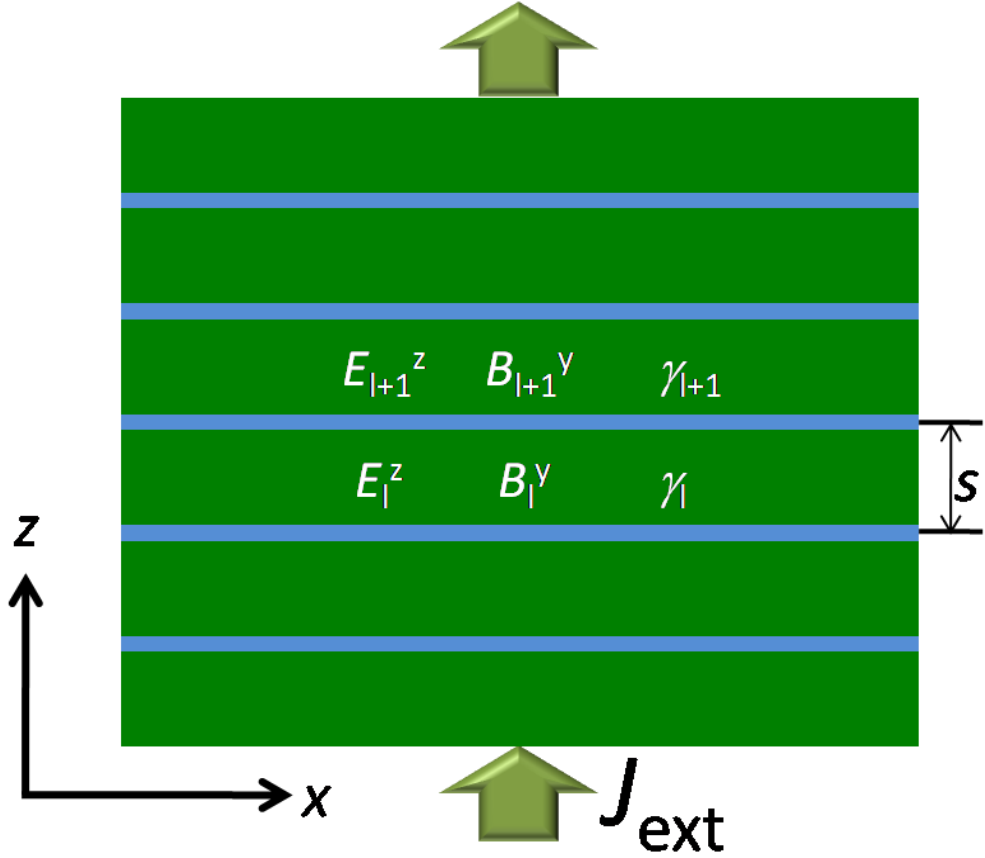

Figure 9. Schematic view of a stack of intrinsic Josephson junctions . The blue/green area denotes superconducting/insulating layers.

Let us concentrate first on the inductive coupling effect since it is the most significant effect in IJJs. A sketch of the model system is shown in Figure 9. For ease of discussion, the thermal effect is neglected and the system is assumed to be infinite along the $y$ direction. Thus the system reduces to two dimensions for the superconductivity phase (electric and magnetic fields are of three dimensions). From the Lawrence-Doniach [77] model for the layered superconductor, one has [78, 79]

$$\nabla \gamma_l = \frac{8\pi^2 \lambda_{ab}^2}{c\phi_0}(\mathbf{J}_{l+1} - \mathbf{J}_l) + \frac{2\pi s}{\phi_0}\mathbf{B}_l \times \hat{\mathbf{z}} \ , \quad (3.1)$$

where $s$ is the period of layered structure, $\gamma_l$ and $\mathbf{B}_l$ the phase difference and magnetic field in the $l$-th junction formed by the $l$-th and $(l+1)$-th superconducting layers, $\mathbf{J}_l$ the current in the $l$-th superconducting layer; the vectors and the gradient operator $\nabla$ refer to the lateral directions. Neglecting the charging effect, the current conservation gives

$$\partial_x\left(J_{l+1}^x - J_l^x\right) + \partial_y\left(J_{l+1}^y - J_l^y\right) = -\frac{1}{s}\Delta_d J_l^z \ , \quad (3.2)$$

where $J_l^z$ is for the current across the $l$-th junction, and

$$\Delta_d Q_l \equiv Q_{l+1} + Q_{l-1} - 2Q_l \quad (3.3)$$

is a second-order difference operator in the $c$ direction. To a good approximation, the magnetic field is taken to be parallel to the $ab$ plane and the electric field be parallel to the $c$ axis in the present system. Then, with Maxwell equations (2.6) and

$$\varepsilon \nabla \cdot \mathbf{E} = 4\pi\rho, \quad (3.4)$$

and Eqs. (3.1) and (3.2), one arrives at

$$\Delta \gamma_l = \left(1 - \frac{\lambda_{ab}^2}{s^2}\Delta_d\right)\left(\frac{8\pi^2 s}{\phi_0 c}J_l^z + \frac{2\pi s\varepsilon}{\phi_0 c}\partial_t E_l^z\right), \quad (3.5)$$

where $\Delta$ is the Laplace operator defined in the lateral directions. Using Eq. (2.7) with the critical current

$$J_c = \frac{c\phi_0}{8\pi^2 \lambda_c^2 s} \quad (3.6)$$

and the ac Josephson relation (2.4), we finally obtain the equation for the phase difference

$$\Delta \gamma_l = \frac{1}{\lambda_c^2}\left(1 - \frac{\lambda_{ab}^2}{s^2}\Delta_d\right)\left(\sin \gamma_l + \frac{4\pi\sigma\lambda_c^2}{c^2}\partial_t \gamma_l + \frac{\lambda_c^2}{c^2/\varepsilon}\partial_t^2 \gamma_l\right). \quad (3.7)$$

Comparing with Eq.(2.8) for a single junction, the lateral space of IJJs is scaled by $\lambda_c$ instead of $\lambda_J$, and the plasma velocity by the light velocity in the material

$$c' = c/\sqrt{\varepsilon} \quad (3.8)$$

instead of the Swihart velocity (2.9). For phase differences non-uniform in the $c$ axis, the term associated with the operator $\Delta_d$ sets the length scale $\lambda_J = \Gamma s$ with $\Gamma = \lambda_c/\lambda_{ab}$. In dimensionless form, Eq. (3.7) reduces to

$$\Delta \gamma_l = (1 - \zeta \Delta_d)(\sin \gamma_l + \beta \partial_t \gamma_l + \partial_t^2 \gamma_l) \quad (3.9)$$

with the inductive coupling constant

$$\zeta = \left(\frac{\lambda_{ab}}{s}\right)^2 \quad (3.10)$$

and the conductivity parameter

$$\beta = \frac{4\pi\lambda_c \sigma}{c\sqrt{\varepsilon}}. \quad (3.11)$$

For BSCCO typically of $\lambda_{ab} = 400$nm, the inductive coupling constant is huge $\zeta \approx 10^5$. The conductivity parameter is typically of $\beta = 0.02$.

Taking away the component generated by the magnetic field associated with the external current in Eq.(2.16), the equations for the phase differences read

$$\Delta\gamma_l = (1-\zeta\Delta_d)(\sin\gamma_l + \beta\partial_t\gamma_l + \partial_t^2\gamma_l - J_{ext}) \quad (3.12)$$

as a good approximation for junction shorter than $\lambda_c$ with $\lambda_c = 200\mu m$ for BSCCO. For the topmost and the bottommost junctions, the equation is to be modified according to the condition determined by the electrodes [80].

It is convenient to write the above equations in a vector form

$$\Delta\boldsymbol{\gamma} = \mathbf{M}\left(\sin\boldsymbol{\gamma} + \partial_t^2\boldsymbol{\gamma} + \beta\partial_t\boldsymbol{\gamma} - J_{ext}\mathbf{I}\right), \quad (3.13)$$

where $(\boldsymbol{\gamma})_l = \gamma_l$, $(\sin\boldsymbol{\gamma})_l = \sin\gamma_l$, $\mathbf{I} = (+1,+1,+1,+1,...)^t$ the unit vector, and the coupling matrix $(\mathbf{M})_{l,l-1} = (\mathbf{M})_{l,l+1} = -\zeta$, $(\mathbf{M})_{l,l} = 1+2\zeta$ and zero otherwise. For a sufficiently thick stack of IJJs, one finds the following two eigenvectors for the coupling matrix $\mathbf{M}$ by inspection

$$\mathbf{M}\mathbf{I}_2 = -(4\zeta+1)\mathbf{I}_2, \quad \mathbf{I}_2 = (+1,-1,+1,-1,+1,-1...)^t, \quad (3.14)$$

$$\mathbf{M}\mathbf{I}_4 = -(2\zeta+1)\mathbf{I}_4, \quad \mathbf{I}_4 = (+1,+1,-1,-1,+1,+1...)^t \quad (3.15)$$

with period 2 and period 4, in addition to the trivial and uniform one $\mathbf{M}\mathbf{I} = \mathbf{I}$.

From Eq. (3.1) and the Maxwell equation (2.6), it is easy to obtain the relation between the phase difference and the magnetic field

$$\nabla\gamma_l = (1-\zeta\Delta_d)\left(\mathbf{B}_l \times \hat{\mathbf{z}}\right). \quad (3.16)$$

Since the superconducting electrodes are as thin as one or two atomic layers, the charging effect is also not negligible in some cases. With the relation [81]

$$\rho_l = -\frac{1}{4\pi\mu^2}\left(A_l^0 + \frac{\phi_0}{2\pi c}\partial_t\varphi_l\right) \quad (3.17)$$

conjugate to Eq.(3.1) for the current, with $\mu$ for the Debye length, the ac Josephson relation (2.4) is modified into

$$\partial_t\gamma_l = \frac{2\pi cs}{\phi_0}(1-\alpha\Delta_d)E_l^z, \quad (3.18)$$

which is conjugate to Eq.(3.16), where

$$\alpha = \varepsilon\mu^2/s^2 \quad (3.19)$$

is the capacitive coupling constant. With the charge conservation relation and Maxwell equations, the equation for the phase differences including the capacitive coupling is derived as [79, 82-84]

$$(1-\alpha\Delta_d)\Delta\gamma_l = (1-\zeta\Delta_d)\left[(1-\alpha\Delta_d)\sin\gamma_l + \beta\partial_t\gamma_l + \partial_t^2\gamma_l\right]. \quad (3.20)$$

For BSCCO, the capacitive coupling $\alpha$ is of order of unity and thus much less significant as compared with the inductive coupling, although it is crucial for the dispersion of longitudinal plasma as will be discussed below.

*III.3 Josephson plasma in IJJs*

Similarly to a single junction, we can look for the Josephson plasma solution for IJJs. For the pure transverse plasma, the dispersion relation is the same as Eq.(2.15) for a single junction, except the length and time scales,

$$\omega^2 = \omega_J^2 + \frac{c^2}{\varepsilon}k_x^2 \quad (3.21)$$

with

$$\omega_J = c/\lambda_c\sqrt{\varepsilon}. \quad (3.22)$$

The Josephson plasma frequency $\omega_J$ has the typical value of 0.5THz for BSCCO. The dispersion relation of the longitudinal Josephson plasma is derived from Eq. (3.20) taking $\beta=0$

$$\omega^2 = \omega_J^2 + \frac{c^2}{\lambda_c^2/\mu^2}k_z^2 \quad (3.23)$$

for small wave number. Without the charging effect, the dispersion relation reduces to the flat one for a single junction as it should be, since the inductive coupling does not play any role for the pure longitudinal wave and the junctions are totally decoupled. The dispersion of the longitudinal plasma is much smaller than the transverse plasma since $\lambda_c \gg \mu$.

Experimentally, the transverse plasma in high-$T_c$ cuprate superconductor was observed by Tamasaku *et al* (1992) [85], and the longitudinal plasma was observed by Tsui *et al* (1994, 1996) [86, 87], Matsuda *et al* (1995) [88] and Kadowaki *et al* (1997) [89]. Josephson plasma provides a useful tool for exploring various properties, such as vortex state, dynamics of quasiparticles, and phase coherence between layers of high-$T_c$ superconductors [90-92].

In IJJs plasma can propagate in intermediate directions. For a stack of $N$ junctions, there are totally $N$ possible wave numbers in the $c$ direction, and a general form of plasma is given as

$$\gamma_l \sim \exp\left(i\frac{n\pi}{N+1}l\right)\exp[i(xk_x - \omega t)], \quad (3.24)$$

with $1 \leq n \leq N$. Substituting into Eq.(3.20), the dispersion relation is given by

$$\omega^2 = 1 + 2\alpha\eta_n + \frac{1+2\alpha\eta_n}{1+2\zeta\eta_n}k_x^2 \quad (3.25)$$

in dimensionless units with

$$\eta_n = 1 - \cos\frac{n\pi}{N+1} \ . \quad (3.26)$$

Therefore, there are *N* typical velocities:

$$c_n^2 = \left(\frac{\omega_n}{k_x}\right)^2 = \frac{1+2\alpha\eta_n}{1+2\zeta\eta_n} \quad (3.27)$$

for $k_x \gg 1$. For large *N*, the maximal velocity approaches the light velocity in the material, which is associated with a wave nearly uniform in the *c*-direction and thus is essentially a pure transverse plasma wave [see Eq.(3.21)]. Notice that the maximal velocity is limited by the total number of junctions in a nonlinear way through the factor $\eta_1$ in Eq.(3.26). The minimal velocity is achieved by the wave with alternating phase in the *c*-direction, which corresponds to the Swihart velocity in a single junction [18].

Peculiar behaviors of flux dynamics emerge because of the coupling between Josephson junctions in the stack. The splitting of transverse plasma into various modes allows solitons to move faster than the plasma, and gives rise to Cherenkov radiation [93]. It was demonstrated that the interaction between the Josephson vortex lattices and plasma produces a photonic band-gap structure [94, 95].

*III.4 Search for THz EM radiation based on IJJs*

The first experimental observation of EM radiation from IJJs is by Kleiner *et al* (1992, 1994) [9, 10] based on BSCCO single crystal. In this experiment, a sample of about 30 $\mu$m in lateral directions and 3 $\mu$m in the stack direction was applied to a *c*-axis bias voltage. The experiment confirmed that the radiation frequency and voltage obey the ac Josephson relation, but the line width of radiation peak is much broader than expected for the case of all junctions synchronized, as shown in Figure 10. This leads the authors to conclude that several groups of tens of junctions are radiating coherently, but the radiations are out of phase among groups. The authors demonstrated that all junctions radiate near the transition temperature $T_c$ [9, 10].

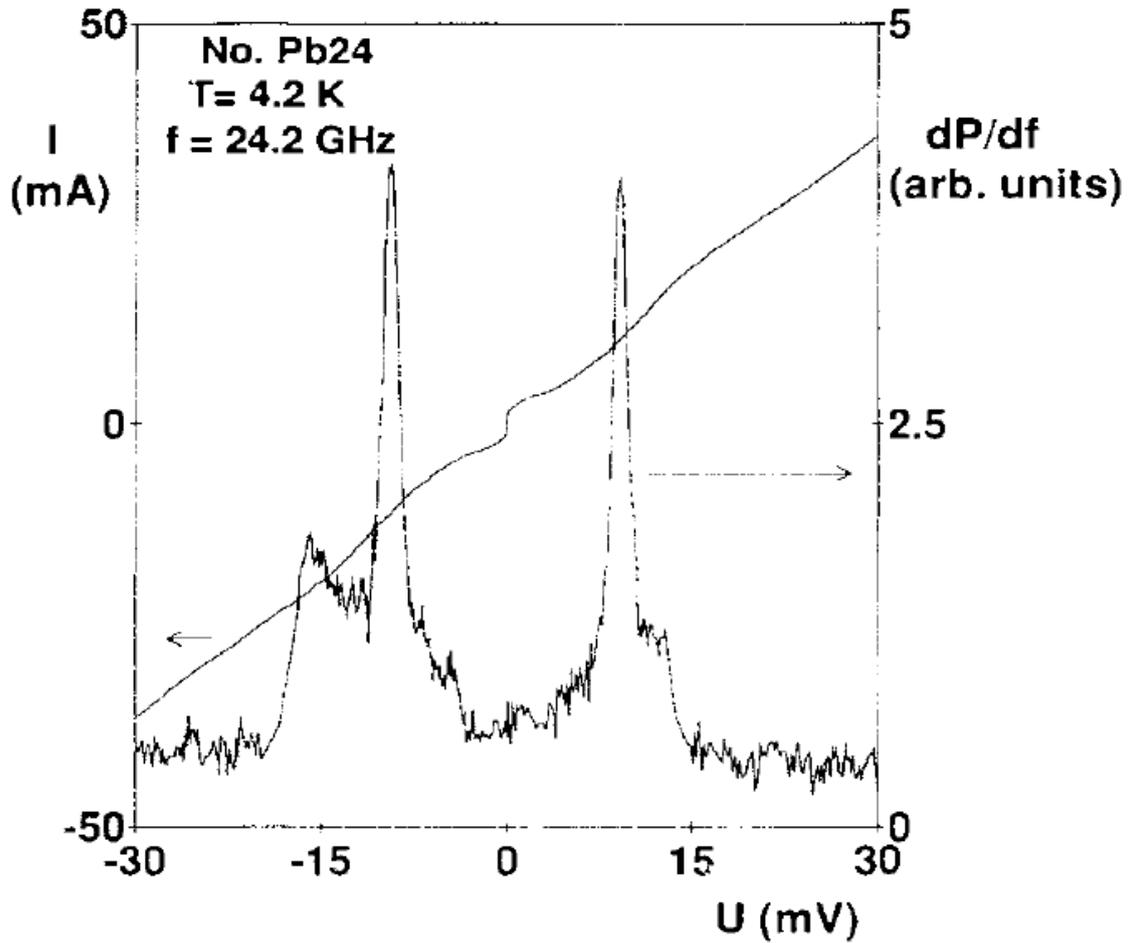

Figure 10. *IV* characteristic (left scale) and emitted microwave power (right scale) of sample $(Bi_{1.785}Pb_{0.125})_2Sr_2CaCu_2O_8$ at T = 4.2 K. The detector frequency is 24.2 GHz. The maximum emission occurring at 9.2mV can be attributed to a packet of 190 junctions. [After R. Kleiner *et al*, Physical Review B 49, 1327-1341 (1994).]

Batov *et al* (2006) [11] reported the radiation from BSCCO in the terahertz band. They used the step-like geometry proposed by Wang *et al* (2001) [96], and the working part is of 3$\mu$m in the lateral directions and consists of about 100 junctions, see Figure 11(a). They used low-noise all-superconducting heterodyne receiver, which was placed 10cm away from the BSCCO. By sweeping the voltage, the radiation was measured at the detector with frequency of ~0.5THz and power in the order of pW, see Figure 11(b). The radiation is very weak because the junctions are not synchronized and the radiation takes place from a single junction.

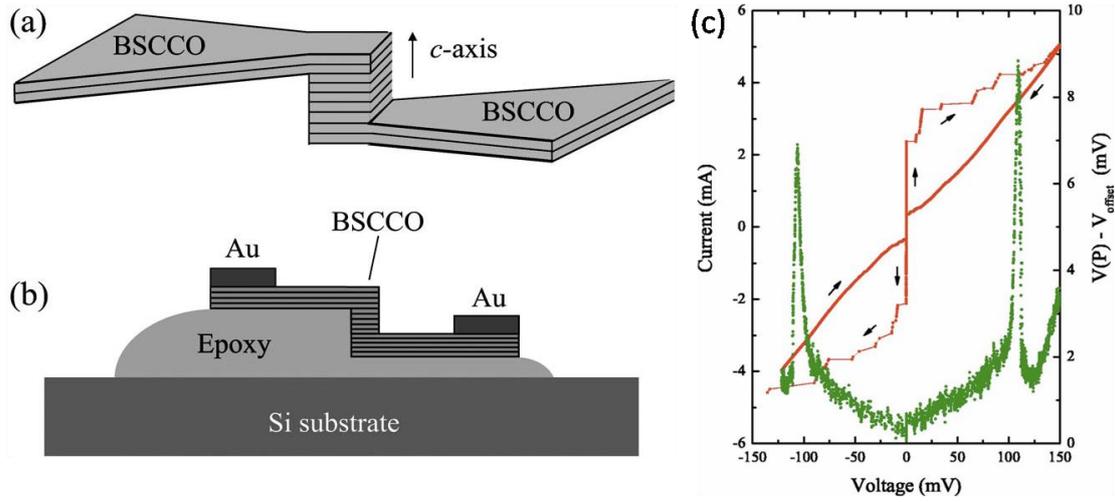

Figure 11. Sketch of the sample layout: (a) BSCCO mesa defined by a double-sided process and (b) cross section of the sample. Dimensions are not to scale. (c) *I-V* characteristics for a specific cycle within the high bias voltage range and the emission signal detected at f =474 GHz for sample 2 at 4.2 K. The lateral size of the stack is 3μm×3μm. [After Batov *et al*, Appl Phys Lett 88, 262504 (2006)].

*III.5 Recent breakthrough in exciting coherent THz radiation*

In 2007, Ozyuzer *et al* (2007) successfully observed a coherent radiation of THz EM waves from BSCCO [14]. Their experimental setup is schematically shown in Figure 12(a). A mesa of about 80 μm and 300 μm in the lateral directions and 1 μm in the stack direction, which corresponds to about 660 IJJ, is fabricated on top of a large substrate of BSCCO single crystal. Radiation was detected by a bolometer located at 20cm away from the mesa, which excludes the possibility of near-field effect. In the high bias region, the *IV* curve becomes back bending caused by the strong self-heating effect [97-104]. Although the sample is immersed in the liquid helium, the temperature of the sample is much higher than the ambient temperature, and almost reaches $T_c$ when the current is large. No radiation is detected in this region other than the black-body, incoherent radiation.

Sweeping back the bias voltage, a jump in the *IV* curve occurs around 0.7V which is accompanied by a sharp peak of power radiation as shown in Figure 12(b). The radiation frequency measured by the spectroscopy is about 0.6THz, which follows the ac Josephson relation when plugging in the bias voltage. It is also demonstrated clearly that the frequency at strong radiation coincides with the fundamental cavity mode of the mesa

$$f = c/2W\sqrt{\varepsilon} \quad (3.28)$$

with *W* the width of junction. The two peaks in the curve of radiation power vs bias voltage are found to correspond to the same radiation frequency, implying that different numbers of resistive (active) junctions are involved at the two voltages. From the detected signals, it is shown that the radiation power is proportional to the number of active junctions squared, which is a direct evidence of coherent and super-radiation. The line width of frequency peak is about 9GHz, which also suggests a coherent radiation since the line width is inversely proportional to the number of junctions in a synchronized oscillation [43]. The maximum radiation power is about $0.5\,\mu\mathrm{W}$.

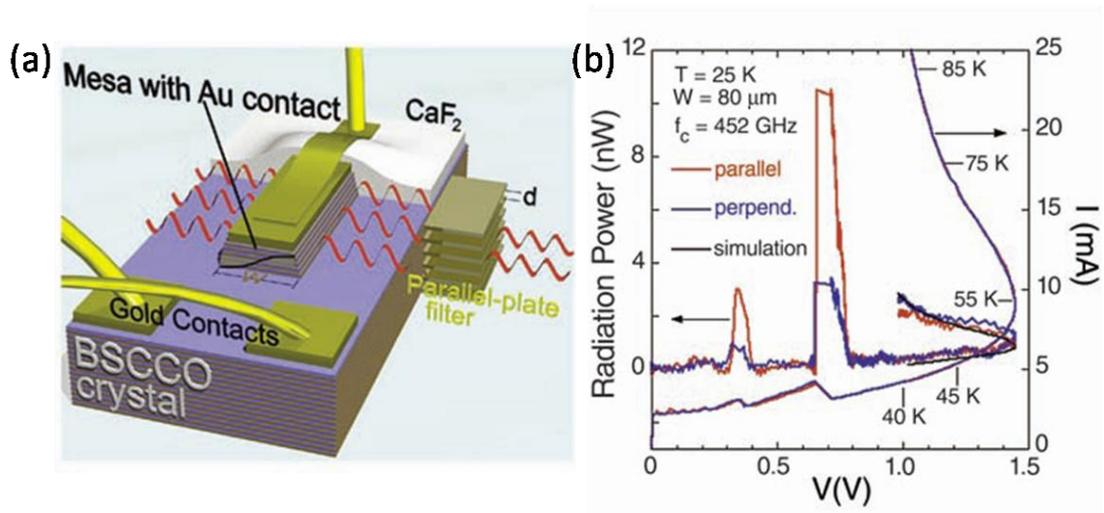

Figure 12. (a) Schematic of the BSCCO mesa. (b) *IV* characteristics and radiation power of the 80μm mesa. The voltage dependence of the current (right vertical axis) and of the radiation power (left vertical axis) at 25 K for parallel and perpendicular settings of the filter with 0.452THz cut-off frequency are shown for decreasing bias in zero applied magnetic field. Polarized Josephson emission occurs near 0.71 and 0.37 V, and unpolarized thermal radiation occurs at higher bias. The black solid line is a simulation of the thermal radiation. [After L. Ozyuzer *et al*, Science 318, 1291-1293 (2007)]

Soon after the breakthrough, Kadowaki *et al* observed a strong radiation as high as 50 $\mu$W with the same sample geometry [105]. A peculiarly anomalous feature appears in the *IV* curve as shown in Figure 13(a). Many frequency harmonics, with the highest frequency up to 2.4THz, are visible at a given voltage due to the strong nonlinear effect. Very recently, stronger radiations by one order of magnitude have been achieved by the same group [106], getting close to the power requested by practical uses. They also measured the far-field radiation pattern, which can shed light on the EM wave and phase dynamics inside the mesa as in the antenna theory.

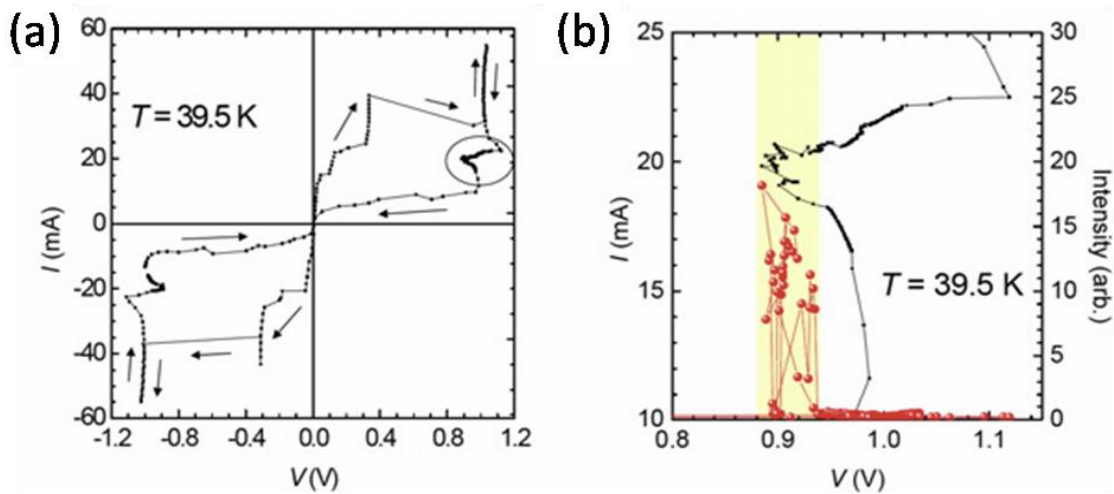

Figure 13. (a) *IV* curve taken at 39.5 K, where the emission is maximal. A circle indicates the region with a sharp voltage dip where strong emission occurs. (b) The region encircled in (a) is shown in an expanded scale. The emitted power intensity is also added as a function of the applied voltage. [After K. Kadowaki *et al*, Physica C 468, 634-639 (2008)]

The samples in these experiments [14] are of a trapezoid shape, hence tend to suppress a perfectly coherent cavity resonance. Recent efforts toward fabrication of rectangle-shaped mesa and excitation of THz emission are reported by Ozyuzer *et al* (2009) [107]. The condition under which the radiation is optimized still needs to be clarified from an experimental point of view.

These encouraging progresses inspire further investigations. Using lower temperature laser microscopy, Wang *et al* (2009) directly visualized the electric field of the junction stack [108]. They observed standing waves of the electric field, thus provided a direct confirmation of cavity modes observed in the previous experiments. The cavity modes along the longer lateral direction are excited in contrast to Ref. [14], see Figure 14. The oscillating electric field seems to be zero at the edges, but it is not very clear due to the attenuation by the gold layer covering the mesa surface. In high bias region where the *IV* curve is back bending caused by self heating, the authors observed that hot spots with temperature higher than $T_c$ are created, and their positions vary with the external current. It is argued that these hot spots can be used to tune the cavity size, and thus the radiation frequency [108].

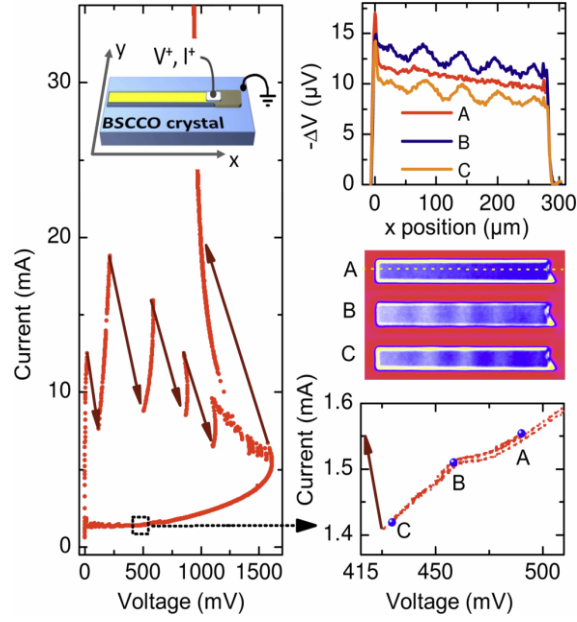

Figure 14. Visualization of standing wave pattern by low temperature scanning laser microscopy (LTSLM). A 40μm× 330μm BSCCO mesa was studied at T= 25 K. Left: current-voltage characteristic (IVC) on large current and voltage scales. Solid arrows denote jumps in the IVC; inset shows the geometry of the device. Right: enlargement of the IVC (bottom) in the region where LTSLM images (A–C; shown above) were recorded. Upper graph shows line scans from images A–C, along the long side of the mesa at half width, cf. dashed line in image A. [After H. B. Wang *et al*, Physical Review Letters 102, 017006 (2009)].

*III.6 Boundary condition for the coupled sine-Gordon equations*

In order to reveal theoretically the mechanism of EM radiation in THz band observed in the recent experiments [14], we need to clarify the boundary condition for the coupled sine-Gordon equations (3.12), or equivalently (3.13). The case of a single junction was investigated by Landenberg *et al* (1965) [35]. Taking the junction as a transmission line connected to vacuum, its impedance is evaluated as [18]

$$Z_J = Z_0 \frac{c}{c'} \frac{d}{W\varepsilon} \sim 10^{-5} Z_0, \quad (3.29)$$

with $W$ the width, $d$ the separation of junction and $Z_0$ the impedance of vacuum. Due to the large impedance of outside space as compared with that inside the junction, the EM waves feels the interface as an open-end circuit, where the current and tangential magnetic field are zero while the electric field achieves its maximum. Since the transmission coefficient at the interface is given by

$$T = \frac{4Z_0 Z_J}{(Z_0 + Z_J)^2} \sim 10^{-4} \ll 1 \quad (3.30)$$

for the EM wave, the junction can be considered as a good cavity approximately.

An analysis on the relation between the electric and magnetic fields at the interface was given by Bulaevskii and Koshelev (2006) for a stack of IJJs [109, 110]. Suppose the right edge of the junctions is at $x=L$ and consider a state uniform along the $y$ direction with the time dependence of EM fields $\exp(-i\omega t)$. At the interface, the magnetic field is related with the electric field by [109, 110]

$$B^y\left(x=+L, k_z, \omega\right) = -\xi\left(k_z, \omega\right) E^z\left(x=+L, k_z, \omega\right) \quad (3.31)$$

where

$$\xi(k_z, \omega) = \begin{cases} k_\omega \sqrt{\varepsilon_d}/\sqrt{k_\omega^2 - k_z^2}, & |k_z| < k_\omega \\ -i k_\omega \sqrt{\varepsilon_d}/\sqrt{k_z^2 - k_\omega^2}, & |k_z| > k_\omega \end{cases} \quad (3.32)$$

with $k_\omega = \omega \sqrt{\varepsilon_d}/c$ and $\varepsilon_d$ the dielectric constant of the dielectric medium attached to the IJJs. One then finds

$$\begin{aligned} B^y\left(x=+L, z, \omega\right) &= -\frac{1}{2\pi} \int dz' E^z\left(x=+L, z', \omega\right) \int \xi\left(k_z, \omega\right) e^{i k_z (z-z')} dk_z \\ &= -\frac{k_\omega \sqrt{\varepsilon_d}}{2} \int dz' E^z\left(x=+L, z', \omega\right) \left[ J_0\left(k_\omega(z-z')\right) + i N_0\left(k_\omega(z-z')\right) \right] \end{aligned} \quad (3.33)$$

where $J_0(u)$ and $N_0(u)$ are the Bessel functions. In experiments [14] $k_\omega L_z \ll 1$ with $L_z$ the height of the mesa, we arrive at

$$B^y\left(x=+L, \omega\right) = -E^z\left(x=+L, \omega\right)/Z, \quad (3.34)$$

with the surface impedance

$$\frac{1}{Z} = \frac{L_z k_\omega \sqrt{\varepsilon_d}}{2} \left(1 - \frac{2i}{\pi} \ln \frac{5.03}{L_z k_\omega}\right) \quad (3.35)$$

when the $z$ dependence of the EM field is neglected since the wave length is much larger than the thickness of the junction stack. For an EM wave of frequency in THz band, $k_\omega \approx 10^{-3} \mu m^{-1}$.

With the continuity condition of EM field at the interface, it is then clear that for a stack of IJJs with $L_z \leq 10 \mu m$ the magnetic field should be much smaller than the electric field at the junction edge, in contrast to the plane wave, and that the EM wave inside the junction stack feels the edge like an open-end circuit. As a good approximation, one can put the tangential magnetic field to zero at the edge of IJJs.

The boundary condition for the phase difference at the edge of junctions is then obtained

$$\partial_x \gamma_l (x=L) = (i\omega) \frac{k_\omega L_z \sqrt{\varepsilon_d}}{2} \gamma_l (x=L) \left( 1 - \frac{2i}{\pi} \ln \frac{5.03}{k_\omega L} \right). \quad (3.36)$$

For a thin junction stack with $k_\omega L_z \ll 1$, it reduces to a Neumann-type boundary condition.

*III.7 Cavity mode caused by modulation of critical current*

In order to achieve a strong emission, a large Josephson plasma oscillation is preferred. Koshelev and Bulaevskii (2008) [111] proposed to introduce spatial modulations of Josephson critical current in IJJ stacks in order to excite cavity resonance. The resonance caused by inhomogeneity in conventional Josephson junctions was discussed decades ago [112, 113]. For a state with phase difference uniform in the $c$ axis of an IJJ stack

$$\gamma = \omega t + \text{Im}\left[ \phi_\omega \exp(-i\omega t) \right], \quad (3.37)$$

Eq. (3.12) is decoupled into

$$\partial_t^2 \gamma + \beta \partial_t \gamma + g(x) \sin \gamma - \partial_x^2 \gamma = 0, \quad (3.38)$$

where $g(x)$ accounts for the spatial modulation of Josephson critical current. Near the cavity mode $\cos(m\pi x/L_x)$ with $m$ an integer, Eq.(3.38) is solved as

$$\phi_\omega = \frac{i g_m}{\omega^2 - k_m^2 + i(\beta + \beta_r)\omega}, \quad (3.39)$$

where

$$g_m = \frac{2}{L_x} \int_0^{L_x} dx \cos\left( \frac{m\pi}{L_x} x \right) g(x), \quad (3.40)$$

and

$$\beta_r = \frac{2 L_z \omega}{\varepsilon_c L_x} \quad (3.41)$$

is the energy dissipation caused by radiation. Since $g_m$ is zero for a uniform critical current as seen in Eq.(3.40), the spatial modulation of the critical current is crucial for the excitation of a large Josephson plasma oscillation in this scheme.

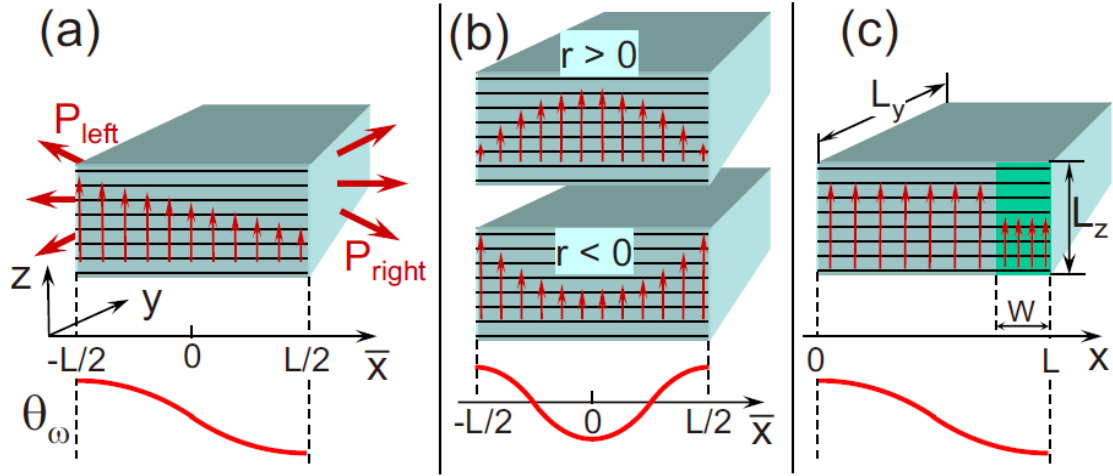

Figure 15. Mesas with various modulations of the Josephson critical current density: (a) linear modulation, (b) parabolic modulation, and (c) steplike suppression of the critical current near the edge. The lower plots illustrate the shapes of lowest excited Fiske-resonance modes. [After A. E. Koshelev and L. N. Bulaevskii, Physical Review B 77, 014530 (2008)]

The dc current is given by

$$J_{ext} = \beta\omega + \langle g(x)\sin\gamma \rangle_{x,t} = \beta\omega + \frac{1}{4}\frac{g_m^2(\beta+\beta_r)\omega}{\left(\omega^2-k_m^2\right)^2+(\beta+\beta_r)^2\omega^2}, \quad (3.42)$$

where $\langle \cdots \rangle_{x,t}$ denotes the average over time and space. The radiation power is given by

$$P_{rad} = \frac{2P_{sc}\omega^2 g_m^2}{\left(\omega^2-k_m^2\right)^2+(\beta+\beta_r)^2\omega}\frac{1}{Z} \quad (3.43)$$

where
$$P_{sc} = L_y L_z \lambda_c J_c E_J / 2 \quad (3.44)$$
and

$$E_J = \frac{\phi_0 \omega_J}{2\pi cs}. \quad (3.45)$$

Several realizations of spatial modulation of the critical current were considered explicitly as in Figure 15, and it is concluded that the radiation power increases significantly at the cavity resonances [111].

Since the modulation of critical current is requested to be identical for several hundred of junctions, some special treatment in sample preparation should be necessary. In Ref. [114], it was discussed that a weak external magnetic field will modulate the critical current, which may be used for the scheme discussed above.

*III.8  π phase kink state*

III.8.1 New solution to the coupled sine-Gordon equations

The experimental observation of the cavity relation of radiation frequency [14] indicates that standing waves of plasma oscillation has been built in the cavity formed by the mesa, which in turn implies that the oscillating part of the phase difference satisfies the Laplace equation. The plasma oscillation should be uniform in the $c$ axis as requested by the observed super-radiation [14]. Taking into account of these properties, the solution to Eq.(3.12) can be generally put as

$$\gamma_l = \omega t + \tilde{\gamma}(x,t) + f_l \gamma^s(x), \quad (3.46)$$

as formulated first by the present authors (2008) [115]. The first term accounts for the finite dc bias voltage according to the ac Josephson relation, and the second for plasma oscillation

$$\tilde{\gamma}(x,t) = A g(x) \sin(\omega t + \varphi) \quad (3.47)$$

with $A$ the amplitude, $g(x)$ an eigenfunction of the Laplace equation satisfying the Neumann boundary condition

$$\partial \gamma / \partial x = 0, \quad (3.48)$$

such as the fundamental mode

$$g_{10}^{\mathrm{r}}(x) = \cos(\pi x / L_x). \quad (3.49)$$

The third term carries the inter-junction coupling via the $l$-dependent factor $f_l$, which has been overlooked in previous studies. We will focus on the fundamental mode $g_{10}^{\mathrm{r}}$ and the generalization to other modes is straightforward.

The general form (3.46) describes a wealth of solutions even giving the constraint imposed by available experimental results. However, the following two solutions in the vector form

$$\boldsymbol{\gamma} = \left[\omega t + A \cos(\pi x / L_x)\sin(\omega t + \varphi)\right]\mathbf{I} + \gamma^s(x)\mathbf{I}_2 \quad (3.50)$$

and

$$\boldsymbol{\gamma} = \left[\omega t + A \cos(\pi x / L_x)\sin(\omega t + \varphi)\right]\mathbf{I} + \gamma^s(x)\mathbf{I}_4 \quad (3.51)$$

diagonalize Eq.(3.12), or equivalently Eq.(3.13) as seen below.

Inserting Eq.(3.50) into Eq.(3.13) and omitting higher harmonics in the Fourier-Bessel expansion (2.48) (note that the solution (3.47) and (3.49) is up to fundamental mode), we arrive at

$$\left[-Ak_x^2 g_{10}^r \sin(\omega t + \varphi)\right]\mathbf{I} + \Delta\gamma^s \mathbf{I}_2 =$$
$$\left[\beta\omega - J_{\text{ext}} + Ag_{10}^r \omega\cos(\omega t + \varphi) - Ag_{10}^r \omega^2 \sin(\omega t + \varphi)\right]\mathbf{I} +$$
$$\left[-J_{-2}\left(Ag_{10}^r\right)\sin(\omega t + 2\varphi) - J_{-1}\left(Ag_{10}^r\right)\sin\varphi + J_0\left(Ag_{10}^r\right)\sin(\omega t)\right]\cos\gamma^s \mathbf{I} +$$
$$\left[J_{-2}\left(Ag_{10}^r\right)\cos(\omega t + 2\varphi) + J_{-1}\left(Ag_{10}^r\right)\cos\varphi + J_0\left(Ag_{10}^r\right)\cos(\omega t)\right]\sin\gamma^s(-4\zeta)\mathbf{I}_2 \quad (3.52)$$

where $k_x = \pi/L_x$. With the orthogonality of the eigenvectors $\mathbf{I}$ and $\mathbf{I}_2$, the coupled sine-Gordon equations is reduced to the following four equations at a given angular frequency [116]:

$$\partial_x^2 \gamma^s = 4\zeta \cos\varphi J_1\left(A\cos(k_x x)\right)\sin\gamma^s \quad , (3.53)$$

$$J_{\text{ext}} = \beta\omega - \frac{\sin\varphi}{L}\int_0^L J_{-1}(Ag_{10}^r)\cos\gamma^s dx \quad , \quad (3.54)$$

$$\frac{A}{2}\beta\omega = \sin\varphi \frac{1}{L}\int_0^L \cos\gamma^s \left[J_0\left(Ag_{10}^r\right) + J_{-2}\left(Ag_{10}^r\right)\right]g_{10}^r dx \quad , (3.55)$$

$$\frac{A}{2}\left(\omega^2 - k_x^2\right) = \cos\varphi \frac{1}{L}\int_0^L \cos\gamma^s \left[J_0\left(Ag_{10}^r\right) - J_{-2}\left(Ag_{10}^r\right)\right]g_{10}^r dx \quad . \quad (3.56)$$

From Eq.(3.53) one derives the $\pi$ phase kink (or generally $(2m+1)\pi$ with $m$ integer) as shown in Figure 16. The width of the kink is given by $\lambda_c/\sqrt{A\zeta} = \Gamma s/\sqrt{A}$ with $\Gamma \equiv \lambda_c/\lambda_{ab}$ the anisotropy parameter, which is the same as an interlayer Josephson vortex when the plasma amplitude is of order of unity. Since a strong inductive coupling is crucial for the phase kink, there is no counterpart in single junctions discussed in Sec. II.

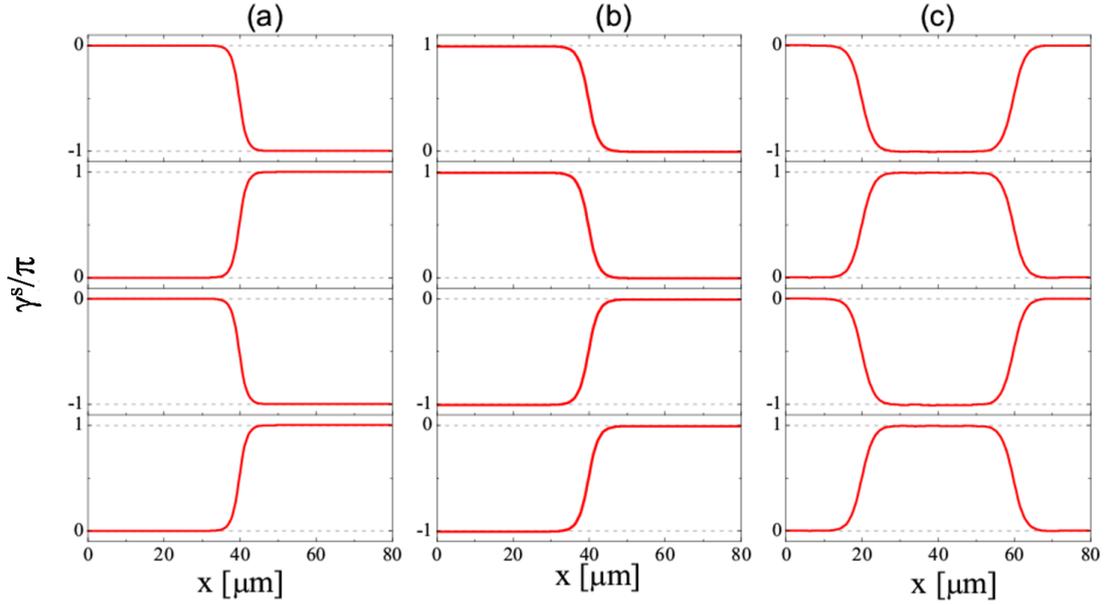

Figure 16. Typical configurations of the static term in the phase difference (3.46): (a) period of 2 layers and (b) period of 4 layers, corresponding to the first current step, and (c) corresponding to the second current step. [After S. Z. Lin and X. Hu, Physical Review Letters 100, 247006 (2008).]

Equation (3.54) is the conservation law of current, with the first term contributed from quasiparticles and the second from the Josephson current. It is clear that the $\pi$ phase kink provides a coupling between the cavity mode of transverse plasma and the uniform dc bias, with which a large dc current can be pumped into the system in the form of supercurrent. With Eq.(3.55) and the recursion relation of Bessel functions $zJ_{\nu-1}(z) + zJ_{\nu+1}(z) = 2\nu J_\nu(z)$, Eq.(3.54) can be simplified to

$$J_{\text{ext}} = \beta\omega\left(1 + A^2/4\right) \quad . (3.57)$$

Namely, the nonlinear *IV* characteristic curve is determined uniquely by the amplitude of plasma oscillation. Solving Eqs.(3.53)~(3.56), one derives the *IV* characteristic curve as depicted in Figure 17.

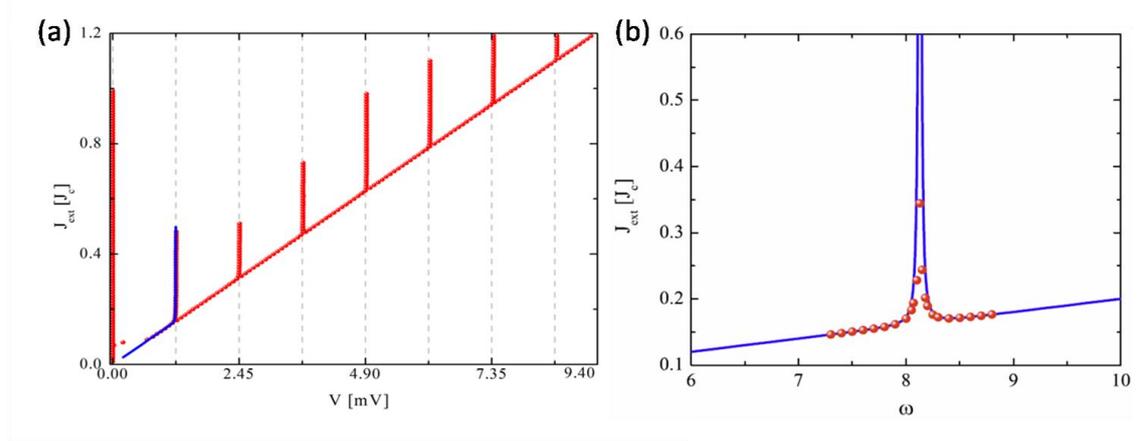

Figure 17. (a) Simulated *IV* characteristics of IJJs in the $\pi$ kink state for $L_x = 0.4\lambda_c$. The vertical dashed lines correspond to the voltage at cavity modes. (b) Theoretical *IV* characteristics around the fundamental cavity resonance at the angle velocity $\omega = \pi/L_x = 7.854$, which corresponds to V= 1.225mV and the frequency approximately f=0.6THz. Red dots are solutions of Eqs.(3.53)-(3.56), and the blue curves are for Eq.(3.58)

When the amplitude is small, one has a simplified expression for the *IV* characteristics as first addressed by Koshelev (2008) [117]

$$J_{ext} = \beta\omega\left[1 + \frac{(I_{10}^r)^2/4}{(\omega^2 - (\pi/L_x)^2)^2 + (\beta\omega)^2}\right] \quad (3.58)$$

with

$$I_{10}^r = \frac{1}{L_x}\int_0^{L_x}\cos(\pi x/L_x)\cos\gamma^s dx \approx 2/\pi, \quad (3.59)$$

where the $\pi$ phase kink is approximated as a step function. The enhancement of the current at the cavity resonance is clearly captured.

It is interesting to remind the following relation for a parallel RLC circuit

$$\langle VI\rangle = \frac{\langle V\rangle^2}{R} + \frac{I_1^2/R}{1/R^2 + (\omega C - 1/L\omega)^2} \quad (3.60)$$

with $I_1$ for the amplitude of oscillating current. The correspondence between the RLC circuit and the coupled sine-Gordon equation (3.13) is clear: $R = 1/\beta$, $C = 1$, and $1/L = (\pi/L_x)^2$.

Comparing Eq.(3.54), or equivalently Eq.(3.59), and Eq.(3.40), it is found that the $\pi$ phase kink plays a similar role as the modulation of critical current. However, the $\pi$ phase kinks are an inherent part of the dynamic state, and evolve automatically in higher cavity modes, in a sharp contrast to the *ad hoc* modulation of critical current.

As revealed above, the phase differences create $\pm\pi$ phase kinks stacked alternatingly along the *c*-axis when they rotate over-all with the angular velocity determined by the bias voltage according to the ac Josephson relation. The $\pi$ phase kinks provide a coupling between the lateral cavity modes of the transverse Josephson plasma and the dc bias voltage. When the bias voltage is tuned to the value corresponding to the cavity frequency of the Josephson plasma, a cavity resonance takes place and the amplitude of plasma is enhanced significantly. The system then carries a large dc Josephson current proportional to the plasma amplitude, since the Josephson current is given by the sinusoidal function of the total phase difference including the rotating phase, the $\pi$ phase kink and the plasma term. Under the dc voltage bias, a large amount of energy is then pumped into the system. As a cartoon picture, the rotating $\pi$ phase kinks work like the pair of pedals of a bicycle to propel energy into the system in the form of standing wave of Josephson plasma oscillation, which then radiates EM wave into space.

It is interesting to evaluate the energy cost of the $\pi$ kink state, as one might worry about whether a state with high energy can be realized [118]. The magnetic energy of a single kink is given by

$$E_{\text{sm}} = \frac{1}{2}\int_0^{L_x} (\partial_x \gamma_l)^2 dx \sim \sqrt{\zeta} \quad (3.61)$$

in units of $\phi_0 J_c / 2\pi c$, since $\partial_x \gamma_l \sim \sqrt{\zeta}$ in a regime of scale $1/\sqrt{\zeta}$ around the center of the $\pi$ kink. However, there appear strong attractive interactions among the alternatingly stacked $\pm\pi$ kinks, which compensate the above self-energy of $\pi$ kink. The total magnetic energy averaged for each junction is given by [119]

$$E_{\text{tm}} = \frac{1}{2N}\int_0^{L_x} \partial_x \boldsymbol{\gamma}^t \mathbf{M}^{-1} \partial_x \boldsymbol{\gamma} dx \sim \frac{1}{\sqrt{\zeta}} \ll 1 \quad (3.62)$$

since $\mathbf{M}^{-1} \sim 1/\zeta$. Therefore, the energy cost of the $\pi$ kink state is negligibly small for a system of strong inductive coupling $\zeta$.

The $\pi$ kink state establishes an EM wave almost perfectly uniform along the *c*-axis despite that the phase difference (3.50) is not uniform. The uniform electric field is trivial since $E_l^z = \partial_t \gamma_l$. For the magnetic field one resorts to Eq.(3.16) and finds $B_l^y = B^y + b_l^y$ with $b_l^y \sim O(1/\sqrt{\zeta})$ at the position of $\pi$ kink and $b_l^y = 0$ elsewhere. This uniform EM wave supports the super-radiation with the power emission proportional to junction number squared.

To be comprehensive it is worthy to notice that besides the two simplest periodic configurations $\mathbf{I}_2$ and $\mathbf{I}_4$ discussed above, there are many other possible kink states as observed by computer simulations. Generally, the value of phase change at the kink is $(2m+1)\pi$ as required by the invariant transformation $x \leftarrow L_x - x$ and $\gamma^s \rightarrow (2m+1)\pi - \gamma^s$ in Eq. (3.53). These $(2m+1)\pi$ kinks give the same coupling as shown in Eq. (3.59) because of the huge inductive coupling. Therefore the kink states occupy finite volumes in the phase space with the same energy, which makes this state easily accessible. Other periodic configurations of $\gamma^s$ are reported in Ref. [120].

In addition to the plasma oscillation, there exists a small oscillation of the center of $\gamma^s$ as well, which produces a spike-like oscillation in electric field [115].

*III.8.2 Radiation power*

The radiation energy depends crucially on the boundary condition, and the radiation feeds back to the dynamic state inside the junctions, which makes an exact treatment of the total system quite difficult. Here, we presume that the energy dissipated inside the junctions is much stronger than the one radiated coherently into the space. This treatment is a perturbative one, and should be checked by the estimate of radiation power thus obtained. We presume that the interface between the junctions and the free space is governed by an effective impedance

$$Z = |Z|\exp(i\theta) \quad (3.63)$$

with $|Z| \gg 1$. Taking into account all harmonics formally and the surface impedance, the *IV* characteristics (3.57) is modified into

$$J_{\text{ext}} = \beta\omega + \frac{1}{4}\beta\omega\sum_{j=1}^{\infty}(jA_j)^2 + \frac{\cos\theta}{L_x|Z|}\omega\sum_{j=1}^{\infty}(jA_j)^2, \quad (3.64)$$

with $A_j$ the amplitude of the *j*-th harmonics. The radiation power can then be estimated by

$$S_r = J_e\omega \bigg/ \left(\frac{\beta|Z|}{2\cos\theta} + \frac{2}{L_x}\right). \quad (3.65)$$

with $J_e \equiv J_{\text{ext}} - \beta\omega$ the excess current pumped into plasma oscillation.

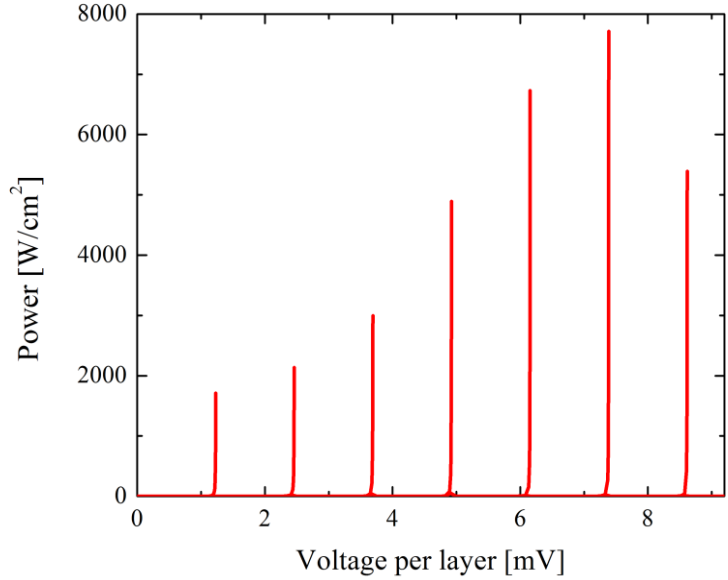

Figure 18. Radiation power versus applied voltage obtained by computer simulation. Sharp radiation peaks occur at all cavity modes.

In Figure 18, we show the simulation results of radiation power estimated by the Poynting vector, which agree well with the estimations by Eq.(3.65) based on the *IV* characteristics shown in Figure 17(a). The maximal radiation power is 8000W/cm$^2$ at the sixth cavity mode. The energy emission at the first cavity mode, which has been achieved in the experiment [14], reaches 2000W/cm$^2$. This corresponds to the total energy of 6mW presuming the same mesa size in the experiment [14], a power sufficient for many practical usages. The optimal efficiency is estimated as

$$Q_\mathrm{e} = \frac{J_\mathrm{e}}{J_\mathrm{ext}} \bigg/ \left( \frac{\beta |Z|}{4\cos\theta} + \frac{1}{L_x} \right) \sim 7.5\% \quad (3.66)$$

at the top of the first current step in Figure 17. Therefore the main part of energy is dissipated as Joule heating in the present system as presumed.

*III.8.3 Other cavity modes*

In the above analysis, the dynamic state is uniform along the $y$ direction, which corresponds to the $(n_x, n_y) = (1,0)$ cavity mode. Modes non-uniform in both lateral directions are also possible. Here we analyze the $(1,1)$ mode as an example. The phase differences for the state of period two layers are described by [116]

$$\gamma = \left[ \omega t + A\cos(\pi x/L_x)\cos(\pi y/L_y)\sin(\omega t + \varphi) \right] \mathbf{I} + \gamma^s(x,y)\mathbf{I}_2 \quad . \quad (3.67)$$

The differential equation of the static phase term $\gamma^s(x,y)$ is

$$\Delta \gamma^s = 4\zeta \cos\varphi J_1\left(A\cos(k_x x)\cos(k_y x)\right)\sin\gamma^s \quad (3.68)$$

with $k_y = \pi/L_y$. Solving the above equation, one obtains the $\pi$ phase kink in two lateral dimensions as displayed in Figure 19. Similar to the (1,0) mode, the 2D $\pi$ phase kink provides a coupling between the uniform dc bias and the (1,1) cavity mode

$$I_{11}^r = \frac{1}{L_x L_y} \int_0^{L_x} \int_0^{L_y} \cos(\pi x/L_x)\cos(\pi y/L_y)\cos\gamma^s dx dy \approx (2/\pi)^2 \quad , \quad (3.69)$$

and a cavity resonance takes place at the frequency

$$\omega = \sqrt{(\pi/L_x)^2 + (\pi/L_y)^2} \quad . \quad (3.70)$$

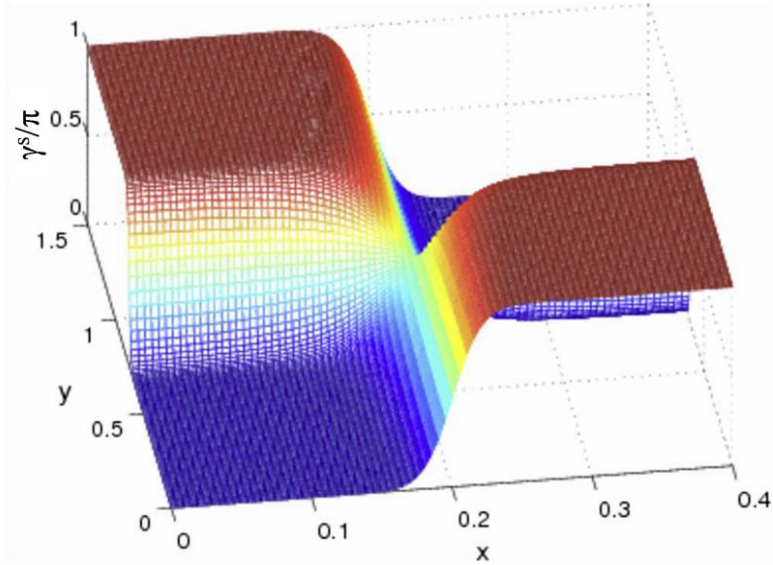

Figure 19. π phase kink for the (1,1) mode in a rectangle mesa. [After X. Hu and S. Z. Lin, Physical Review B 78, 134510 (2008)].

*III.8.4 Radiation pattern*

Here we calculate the far-field radiation pattern from the stack of Josephson junctions. For this purpose, we resort to the Huygens principle in which the pattern is determined by the oscillation of the EM fields at the edges of samples, which can be casted into the edge magnetic current and electric current in the formula of equivalence principle [121]. Since there exists a significant impedance mismatch, the electric current produced by the oscillating magnetic field is much smaller than the magnetic current produced by the oscillating electric field, so we can neglect the contribution from the electric current. The equivalent magnetic current in the dimensionless

units is given by

$$\mathbf{M}_e = \mathbf{E}_e \times \mathbf{n} \quad, \quad (3.71)$$

where $\mathbf{E}_e$ is the oscillating electric field at the edges of mesa and the unit vector $\mathbf{n}$ is normal to the edges. Since the radiation pattern critically depends on the geometry of the source, we need to consider the 3D system. The coordinates for the 3D system are sketched in Figure 20(a). We use the similar dimension as in the experiment [14], namely $L_x = 80\mu\text{m}$, $L_y = 300\mu\text{m}$ and $L_z = 1\mu\text{m}$. Since the thickness of mesa is much smaller than the wavelength of the EM wave, the sources can be treated as uniform along the $c$-axis. In this case, the far-field Poynting vector in the dimensionless units is

$$\mathbf{S}_r = \frac{\omega^2 L_z^2}{32\pi^2 r^2 \varepsilon_d^{3/2}} |\mathbf{G}|^2 \mathbf{e}_r \quad (3.72)$$

with

$$\mathbf{G} = \oint_{\text{edges}} M_e(r') \exp\left(-i\frac{\omega}{\sqrt{\varepsilon_d}} \mathbf{r}' \cdot \mathbf{e}_r\right) (\mathbf{e}_r \times \mathbf{e}_{r'}) dl' \quad, (3.73)$$

where the integral is taken over the perimeter of the crystal [111, 122]. The interference mainly comes from the $y$ coordinate of the source since the corresponding dimension is comparable to the wavelength.

Since the oscillating electric field in the cavity mode $(n_x, n_y)$ is given by

$$E^z = A\omega \cos(n_x \pi / L_x) \cos(n_y \pi / L_y) \quad (3.74)$$

in the frequency domain, the radiation pattern can be calculated directly using Eqs.(3.71)-(3.73), which has been reported in literatures [111, 122]. At cavity resonances, components in the oscillating electric field associated with higher harmonics are not negligible, and should also be included into the calculation of far-field radiation pattern. The results at current steps corresponding to the cavity modes (1,0) and (1,1) are shown in Figure 20(b) and (c). For mode (1,0), the radiation power is maximal at the top of the mesa and at the middle of short side, while takes the minimum at the middle of long side. For mode (1,1), the radiation power is minimal at the top of the mesa and at the middle of both the short and long sides.

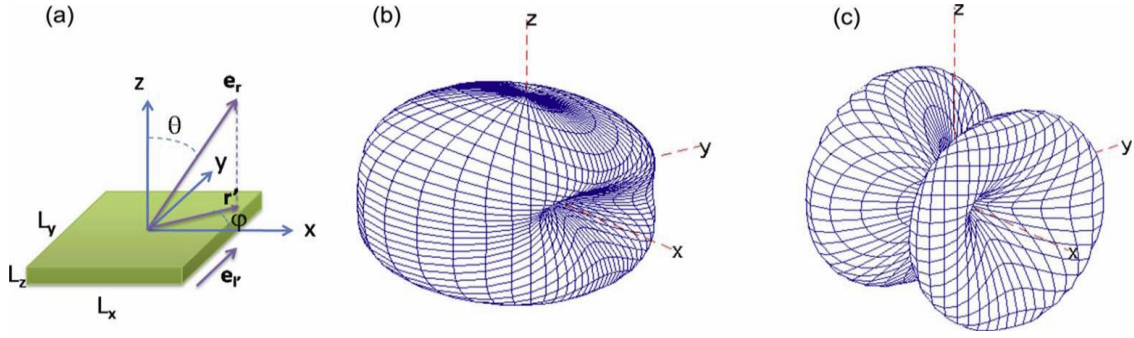

Figure 20. (a) Coordinate system for the calculation of radiation pattern from the mesa. (b) Radiation pattern of the mode (1,0). (c) Radiation pattern of the mode (1,1). Here $L_x$=80 μm, $L_y$=300 μm, and $L_z$=1 μm. [after S. Z. Lin and X. Hu, Physical Review B 79, 104507 (2009)].

*III.9 Cylindrical and annular geometry*

In a rectangular mesa, the spatial part of the plasma term of the lowest cavity mode satisfying the right boundary condition is $\cos(\pi x/L_x)$. Unfortunately, both $\cos(\pi x/L_x)$ and $\sin(\pi x/L_x)$ are eigen functions of the Laplace equation, and give the same wave number $\pi/L_x$. Therefore, the dependence of frequency on the system size cannot determine uniquely the mode. Cylindrical geometry was proposed to identify uniquely the EM mode inside the junctions [116, 123]. For the cylinder geometry, the radial part of eigen function of the Laplace equation is given by the Bessel function. The boundary condition for the cylinder geometry determines uniquely the wave number, and vice versa, since the zeros of the Bessel functions are different from the zeros of their derivatives, contrasting to sine and cosine functions in the rectangle geometry. Therefore, measuring by experiments the frequency of EM radiation from a cylindrical mesa of given radius and assigning the mode by the properties of Bessel functions enable one to identify uniquely the right boundary condition for the EM waves. From the boundary condition suitable for thin mesas revealed theoretically, the frequency of radiation should be given by the inverse of radius with the coefficient equal to zero of derivative of Bessel function.

The spatial part of the $(m,n)$ cavity mode in a cylindrical sample is given by

$$g_{mn}^c(\mathbf{r}) = J_m\left(\chi_{mn}^c \rho/a\right)\cos(m\phi) \quad (3.75)$$

with the cylindrical coordinate $\mathbf{r}=(\rho,\phi)$, $a$ the radius of the cylinder, and $\chi_{mn}^c$ the $n$-th zero of derivative of Bessel function $J_m(z)$. The lowest four zeros of derivative of Bessel functions are given by $\chi_{11}^c=1.8412$, $\chi_{21}^c=3.0542$, $\chi_{01}^c=3.8317$ and $\chi_{12}^c=5.3314$. The $\pi$ phase kinks associated with the four lowest cavity modes are displayed in Figure 21.

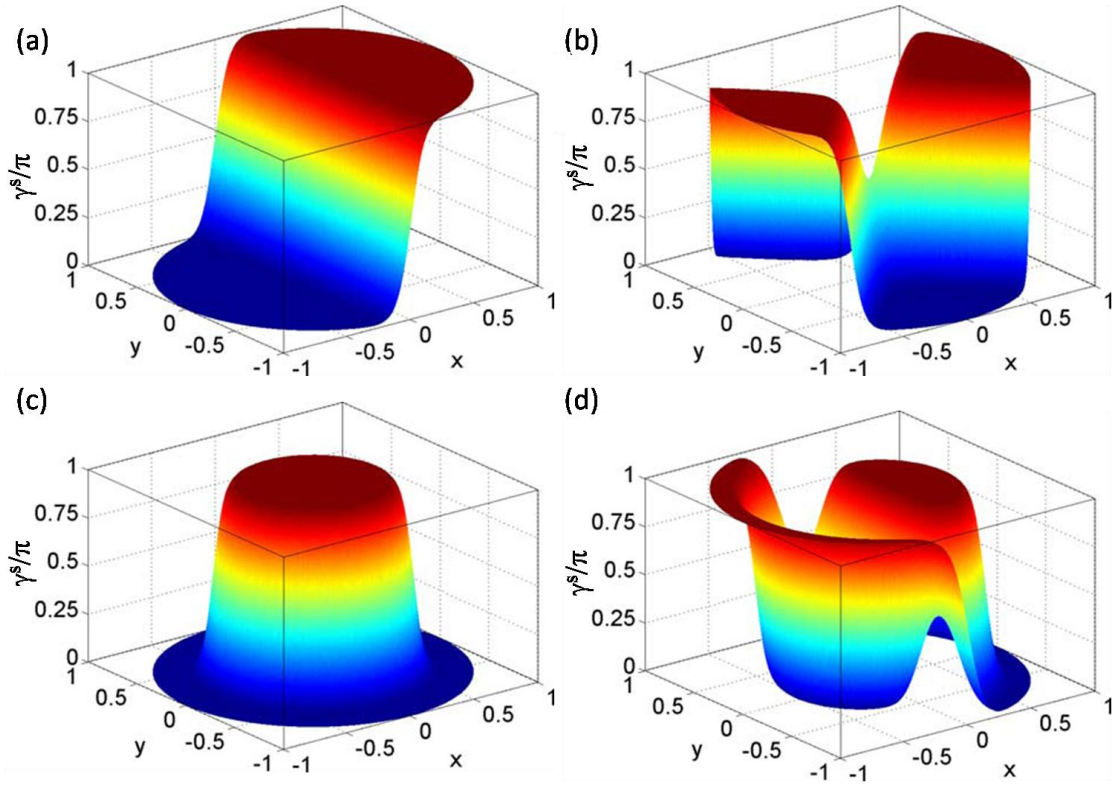

Figure 21. Spatial distribution of the static phase kink for (a): (1, 1), (b): (2, 1), (c): (0, 1) and (d): (1, 2) mode of a cylindrical mesa. [After X. Hu, and S. Z. Lin, Physical Review B 80, 064516 (2009).]

In a recent experiment based on cylindrical mesa [124], the wave number of cavity mode was given by $k \approx 1.87/a$. Since it is very close to the (1,1) mode, the experiment revealed that the electric field presumes maximal at the edge of sample, and the tangential component of the magnetic field is approximately zero, which corresponds to the Neumann-type boundary condition for the phase difference.

A hurdle for the present technique to excite EM waves of high frequency is the heating effect, since the corresponding higher dc voltage injects large currents into the sample resulting in severe Joule heating. One way to overcome this effect may be to dig a hole in the superconductor mesa, rendering for example a cylindrical one to annular, which reduces the cross section, and thus the total current and Joule heating. The inner surface of an annular mesa may help leaking heat generated in the mesa additionally.

For the annular geometry, the spatial part of the plasma term is given by

$$g_{mn}^{a}(\mathbf{r}) = \left[ J_m\left(\chi_{mn}^{a}\rho/a_o\right) + pN_m\left(\chi_{mn}^{a}\rho/a_o\right)\right]\cos(m\phi) \quad , (3.76)$$

where $N_m(z)$ is the Bessel function of second kind and $a_o$ is the outer radius. The Neumann boundary condition should be satisfied at both the outer and inner edges, which determines the coefficient $\chi_{mn}^{a}$

$$J_m'\left(\chi_{mn}^{a}a_i/a_o\right)N_m'\left(\chi_{mn}^{a}\right) - J_m'\left(\chi_{mn}^{a}\right)N_m'\left(\chi_{mn}^{a}a_i/a_o\right) = 0 \quad (3.77)$$

as a function of the aspect ratio $a_i/a_o$, where $J_m'(z)$ and $N_m'(z)$ are the first derivatives, and $a_i$ is the inner radius. The coefficient $p$ is then given by

$$p = -J_m'\left(\chi_{mn}^{a}\right)/N_m'\left(\chi_{mn}^{a}\right) \quad . (3.78)$$

The dependence of wave number on the aspect ratio $a_i/a_o$ is depicted in Figure 22. For (1,1) and (2,1) modes, the wave number decreases with increasing ratio $a_i/a_o$, whereas an opposite trend is seen for the (0,1), and a non-monotonic behavior is found for the (1,2) mode, which can be understood from the redistribution of magnetic field when a hole is dug in the cylindrical sample. Therefore, tailoring the sample shape can be a fruitful way to tune the radiation frequency.

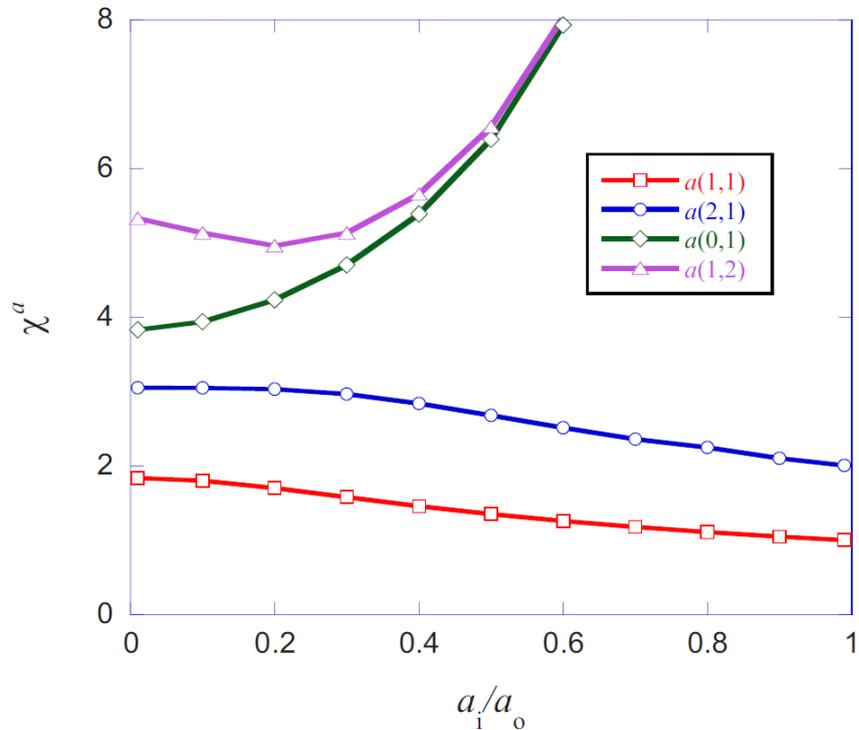

Figure 22. Wave number as a function of the radius ratio $a_i/a_o$ for the lowest four modes for the annular geometry. [After X. Hu, and S. Z. Lin, Physical Review B 80, 064516 (2009).]

*III.10 Other possible solutions*

*III.10.1 McCumber state*

Equations (3.20) have solutions uniform along the $c$ axis. However, since the system is decoupled into identical single junctions in these solutions, some scheme is requested to stabilize the state. We discuss two of these solutions below.

Bulaevskii and Koshelev (2007) proposed to use a short but thick stack of IJJs to generate THz wave [125], in which the synchronization of the phase dynamics is realized by the feed-back effect of radiation. The IJJs are sandwiched by two gold electrodes, which prevent the interference of EM waves radiated from two sides. For stack thickness comparable to the wavelength of EM in the dielectric medium coupled to the junctions, the impedance mismatch is small, according to Eq.(3.35). Under the boundary condition of an effective impedance $Z$, the decoupled equations of (3.38) is solved for $g(x)=1$ to give $\phi_\omega$ defined in Eq. (3.37)

$$\phi_\omega = -\frac{1}{\omega^2 + i\beta\omega} + \frac{i\cos(\omega x)/(\omega + i\beta)}{Z\omega\sin(\omega L_x/2) + i\omega\cos(\omega L_x/2)}. \quad (3.79)$$

The first term represents the Josephson oscillation, same as Eq.(2.32), and the second term corresponds to the propagating waves due to radiation. In the limit of $\omega L_x \ll 1$, the total radiation power is given approximately in a closed form by

$$\frac{P_{\text{rad}}}{L_y} = \frac{\phi_0^2 \omega_J^4 N^2}{64\pi^3 c^2 \omega} R(r) \quad (3.80)$$

with

$$R(r) = \frac{r^2 \varepsilon_c^2}{(r\varepsilon_c + L_\omega)^2 + 1}, \quad (3.81)$$

where $r = L_x/L_z$ and

$$L_\omega = \frac{2}{\pi}\ln\left[\frac{5.03}{k_\omega L_z}\right]. \quad (3.82)$$

The total radiation power grows as $N$ squared, indicating a super-radiation. The radiation modifies the *IV* curve into

$$J_{\text{ext}} = \beta\omega + \frac{1}{2}\langle\text{Im}[\phi_\omega]\rangle_x = \beta\omega + \frac{\beta}{2\omega^3} + \frac{R(r)}{2\omega^2 \varepsilon_c r} \quad (3.83)$$

where the first term accounts for the quasiparticle channel, the second for the plasma oscillation and the last term for the radiation. The current is enhanced by the radiation.

Based on these considerations, the authors proposed optimal sizes of the stack of IJJs for radiation $L_z = 40\mu$m and $L_x = 4\mu$m with $L_y > 300\mu$m. The total radiation power is about $P_{rad}/L_y = 30$mW/cm with Joule heating 15W/cm$^2$. The stability of the solution was also discussed, and it is found that the radiation and dissipation contribute to stabilizing the state.
It is noticed that in the region of $\omega > 1$ only a small amount of power can be pumped into the plasma oscillation according to Eq.(3.79), which limits the radiation power. On the other hand, the maximum efficient at frequency 0.75THz is as high as 60%, due to the relatively small surface impedance.

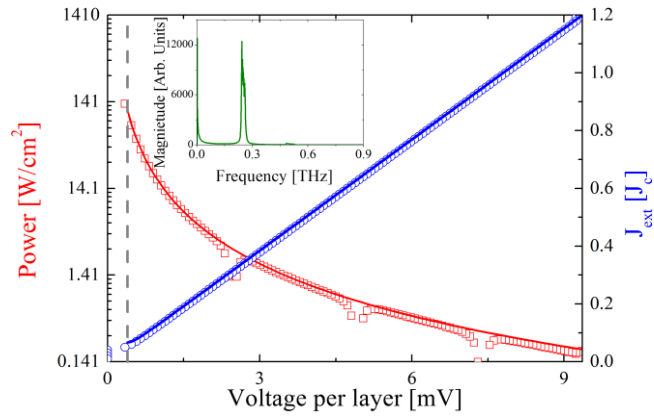

Figure 23. Radiation power from the McCumber state and its corresponding *IV* characteristics. Symbols are for simulations with radiating boundary condition and lines are for theory. The vertical dashed line is the re-trapping point obtained from the theory. The inset is the frequency spectrum at the strongest radiation. The results are obtained with the impedance $|Z| \approx 1000$. [After S. Z. Lin and X. Hu, Physical Review B 79, 104507 (2009)]

The radiation from the McCumber state in a sample with dimensions similar to that in the experiment [14] was studied by the present authors [119]. The *IV* curve and radiation power are shown in Figure 23. The *IV* curve is almost linear since the plasma oscillation is always weak. The dependence of radiation power on voltage is almost monotonic and reach maximum near the re-trapping point, where the input power is insufficient for a phase particle to travel through the tilted washboard potential. Due to the weak plasma oscillation, no harmonics is visible in the frequency spectrum as shown in the inset of Figure 23. The local minima in the radiation power are caused by the small spatial modulation of EM fields, which changes with the voltage.
The re-trapping current can be estimated as the minimal current in Eq.(3.83), and in the weak damping limit $\beta \ll 1$, it is given by

$$J_\mathrm{r} = \frac{4}{3}\beta^{3/4}\left(\frac{3\cos\theta}{|Z|L_x} + \frac{3\beta}{2}\right)^{1/4} . \quad (3.84)$$

Recently Koyama *et al* (2009) [126] performed computer simulations on the phase dynamics and EM fields of IJJs immersed in vacuum, solving both the coupled sine-Gordon equation inside the IJJs and Maxwell equations in the vacuum with a boundary condition of perfect absorption at the outer boundary of vacuum. The advantage of their approach is that the interference of EM waves in space and the feed-back to the phase dynamics in IJJs can be taken into account. To simplify the calculation, they take the phase difference uniform along the *c*-axis. This makes their solution similar to the McCumber state. They also calculated the far-field radiation pattern and found a one similar to a dipole antenna.

It is well known that the McCumber state of a long single junction is unstable near the cavity resonance and the system evolves into soliton state [127]. The stability of the McCumber state in a stack of IJJs is investigated by Nonomura (2009) [128]. While the state is stable when the impedance mismatch is small, consistent with the analysis in Ref. [125], it becomes unstable with large surface impedance mismatch, and evolves into the $\pi$ kink state when small perturbations are added [119, 129].

*III.10.2 Another standing wave solution*
Very recently, Tachiki *et al* (2009) proposed another state in an IJJ stack based on computer simulations [118]. The authors found a standing wave of Josephson plasma which is uniform along the $c$ axis. At the voltages
$V = 2m\pi/L_x$ (3.85)
with $m$ integers, there appear resonating peaks in the *IV* characteristics, where the amplitude of Josephson plasma is enhanced and strong emissions are evaluated at the edges, see Figure 24.

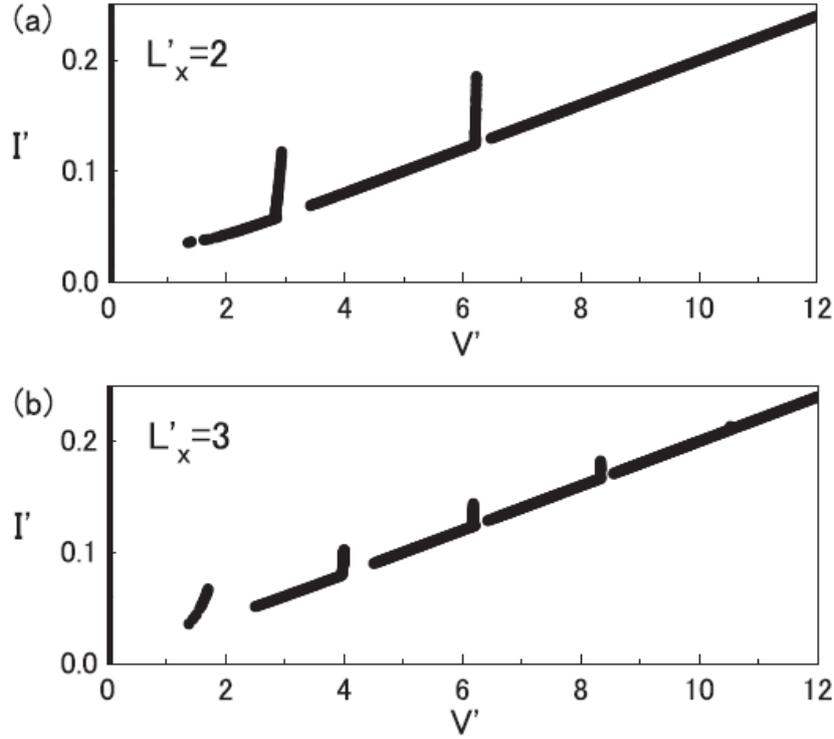

Figure 24. *IV* curve for (a): L=2 and (b): L=3 obtained by computer simulation. [After M. Tachiki *et al*, Physical Review Letters 102, 127002 (2009)].

The junction-width dependence of voltage at which a strong radiation occurs is the same as the zero-field Fiske steps caused by motions of solitons in a single junction [130], and is different from the experimental observation [14]. In a relatively short junction and with slightly open boundary condition, the soliton becomes incomplete, and its shuttle motions cause phase dynamics resembling to a standing wave [131]. It is well known that due to the strong mutual repulsion solitons tend to form triangular lattice, which cannot support coherent radiation. The stability of the state proposed in Ref. [118] remains a matter of discussion [130].

**IV. Radiation under finite magnetic field**

As discussed in Sec. II.5 for a single junction, one can stimulate transverse Josephson plasma and radiation of EM wave by driving the Josephson vortices induced by an in-plane magnetic field. However, for IJJs a triangular lattice of Josephson vortices only generates an EM oscillations out-of-phase in neighboring junctions. Therefore, in order to get strong radiation, one has to control the vortex configuration and achieve a rectangular Josephson vortex lattice (JVL). In long junctions the stable JVL is triangular due to the repulsive interaction. For short junctions, the surface effect may win which tends to align the vortices and thus favors rectangle lattice [132-137]. Besides the surface effect, it is also suggested that pancake vortices can render rectangular JVL[138].

The JVL experiences structure transitions when it is driven by a $c$-axis current. In Ref. [139], several different lattice structures are observed associated with different characteristic velocities $c_n$'s, see Figure 25.

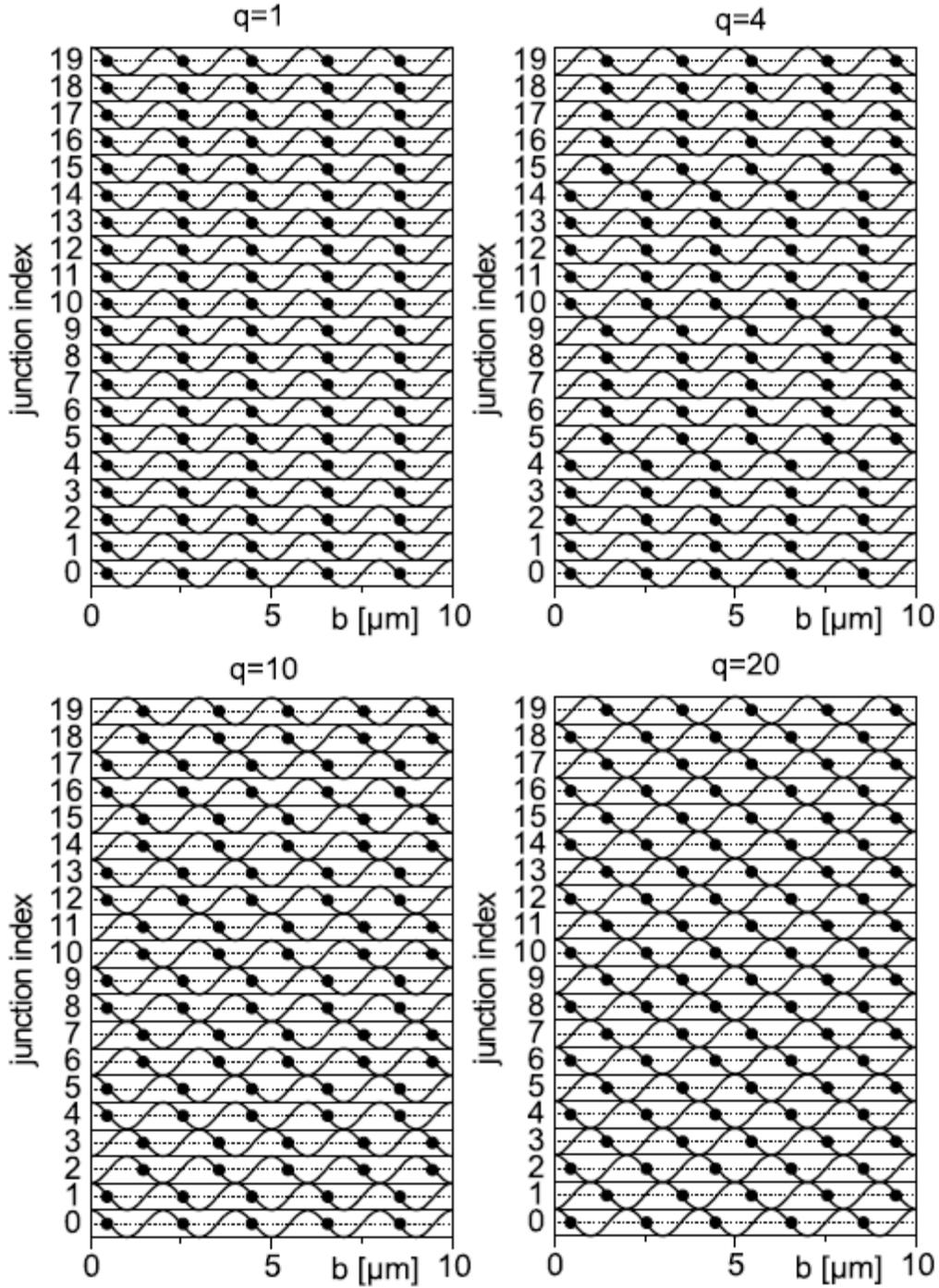

Figure 25. Configurations of Josephson vortices at different characteristic plasma velocities. The black dots indicate the core of vortices. [After R. Kleiner *et al*, Physica C 362, 29-37 (2001)].

The rectangular JVL corresponds to the $c_1$ mode and is realized by relatively large bias. The stability of this vortex configuration is under intensive investigation [79, 140-142]. In Refs. [79, 140], vortex configurations in infinitely long junctions are studied when the magnetic field and external current are tuned. It is revealed that there are two major stable branches, with the lower velocity branches close to triangular configuration and the higher velocity branches close to rectangular one. The lattice structure is selected by the boundary condition in the *c* axis. To date, there still lacks of a general protocol to realize the in-phase, rectangular JVL.

*IV.1 Experiments*

There are several experimental attempts to stimulate THz radiation under magnetic field. Kadowaki *et al* (2006) used IJJs of underdoped BSCCO single crystal with the sample size $L_y = 5\mu m$ and $L_x = 10\mu m$ in the direction parallel and perpendicular to the applied magnetic field, respectively, and thickness $L_z = 0.9\mu m$ [143]. The radiation is detected by a bolometer, and it is found that the radiation from the edge where the Josephson vortices are moving towards is 3-5% higher than that from the opposite edge. The total radiation power is estimated to be $100\,\text{W}/\text{cm}^2$.

Bae *et al* (2007) performed an advanced experiment in which they integrated the generator and detector both made of IJJs on the same substrate [144]. In the generator there are about 30 IJJs and $L_x \approx 10\mu m$. The radiation is detected by monitoring the *IV* response of the IJJs in the detector. A schematic view is shown in Figure 26(a). The *IV* curve for the generator IJJs is hysteretic and several different behaviors are spotted, as shown in Figure 26(b). When the bias current is low, the Josephson vortices are pinned. In the intermediate bias, Josephson vortex flow branches resulted from phase locking of JVL to plasma modes are observed. Comparing the terminating voltage and the maximum velocity of vortex flow, the authors conclude that the outmost branch corresponds to the $c_1$ mode. In high bias region, multiple quasiparticle branches appear. These complicated behaviors in *IV* characteristics are not fully understood and are a subject of intensive study [145]. Measuring the Shapiro steps in the detector IJJs, the author detected a radiation power ~10nW from the generator in the Josephson vortex flow branches. The radiation frequency is about 1THz.

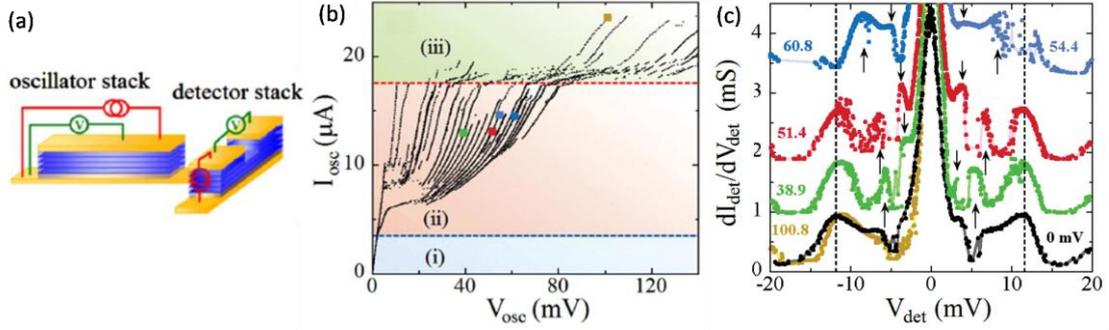

Figure 26. (a) Schematic of sample and measurement configurations, showing the oscillator stack (left) and the detector stack (right) of intrinsic Josephson junctions. (b) Collective JVF multiple sub-branches and bias voltages $V_{osc}$ (solid squares) of the oscillator stack for H = 4 T at T= 4.2K. The contact resistance was numerically subtracted. (c) Responses of the detector stack revealed in its differential conductance corresponding to the biases of the oscillator in (b). Each curve is shifted vertically for clarity. [After M. Bae *et al*, Physical Review Letters 98, 027002 (2007)].

*IV.2 Theoretical studies*

Machida *et al* (2001) calculated the radiation power by numerical simulations [146]. Several plasma modes were observed and the in-phase mode is attained at intermediate current, where a strong THz emission is expected. Later, Tachiki *et al* (2005) performed large-scale computer simulations using the state-of-art supercomputer on a system of several hundred junctions [82]. The system is of $L_x = 100\mu m$ in the direction perpendicular to the magnetic field, and is presumed to be uniform along the direction of external magnetic field. They solved Eq.(3.20) for the junctions and Maxwell equations inside the dielectric medium attached to the junctions. While the configuration of the Josephson vortices is distorted slightly from the rectangular configuration, the radiation power reaches $6000W/cm^2$.

Bulaevskii and Koshelev (2006) [110] investigated the flow of Josephson vortices under the dynamic boundary condition (3.36). They showed that the radiation power caused by the motion of rectangular vortex configuration is about $1500W/cm^2$.

Careful numerical simulations on a 2D system of 30 junctions with $L_x = 30\mu m$ have been carried out recently [147]. Equations (3.20) were solved in the presence of magnetic field under the dynamic boundary condition with the plane-wave approximation. It is found that a rectangular JVL is realized when the driving current is high. Fiske modes are then excited similarly to the case of a single junction, with the strong radiation peak appearing when the velocity of the Josephson vortices matches the velocity of plasma $c_1$, or equivalently the separation of Josephson vortices equals to the wavelength of Josephson plasma. The radiation power is about $500\,\mathrm{W}/\mathrm{cm}^2$, and the radiation frequency can be tuned almost continuously by the applied magnetic field and bias voltage, as shown in Figure 27(a). The configuration of Josephson vortices is overall rectangle except some sliding motions of vortices caused by the strong Josephson plasma, as shown in Figure 27(b).

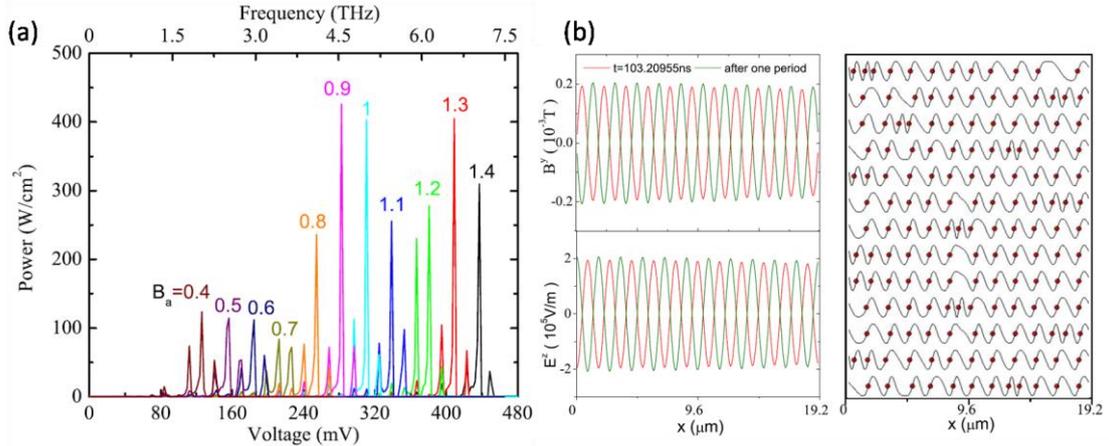

Figure 27. (a) Radiation power at different applied magnetic fields in units of Tesla as a function of the voltage across the stack of 30 IJJs. The frequency in the upper axis is determined by the ac Josephson relation. (b) Left: snapshots for the magnetic and electric fields at the largest resonance for $B_a$=1T, where the dc electric fields has been subtracted. Right: snapshot of vortex configuration. [After S. Z. Lin, X. Hu, and M. Tachiki, Physical Review B 77, 014507 (2008).]

**V Non-Josephson radiations**

*V.1 Cherenkov radiation*

So far we have focused on the EM radiation based on the Josephson effects. There exist other possibilities to excite THz wave in high-$T_c$ superconductors. In the following, we briefly review the Cherenkov radiation, radiation by quasiparticle injection and radiation by moving Abrikosov vortices in high-$T_c$ superconductors.

Since the coupled sine-Gordon equations are Lorentz invariant, relativistic behaviors emerge in Josephson junctions. In a conventional junction, the maximum velocity of soliton and Josephson vortex is limited by the Swihart velocity, and no Cherenkov radiation is expected. However, when the phase dynamics becomes nonlocal [148, 149], i.e. Josephson penetration depth is larger than London penetration depth, the velocity of soliton can exceed the plasma velocity and Cherenkov radiation occurs [150]. In IJJs, the dispersion relation of Josephson plasma splits into $N$ branches, as discussed in Sec II, which allows for faster motion of soliton than plasma velocity.

Cherenkov radiations in artificial Josephson junctions have been investigated both theoretically and experimentally for a long time [151-153]. The Cherenkov radiation from BSCCO is first reported experimentally by Hechtfischer *et al* (1997) [93]. They used a stack of about 100 IJJs of lateral size of $20\mu m \times 20\mu m$, and observed Josephson radiation in the lower magnetic field region. When the magnetic field is increased, a broadband radiation with two orders of magnitude larger power is detected at frequency ranging from 7 to 16GHz as shown in Figure 28. The radiation frequency is two orders of magnitude smaller than that of Josephson radiation. The authors argued that the Cherenkov radiation accounts for the observations with low frequencies. This was further supported by numerical simulations on inductively coupled sine-Gordon equations.

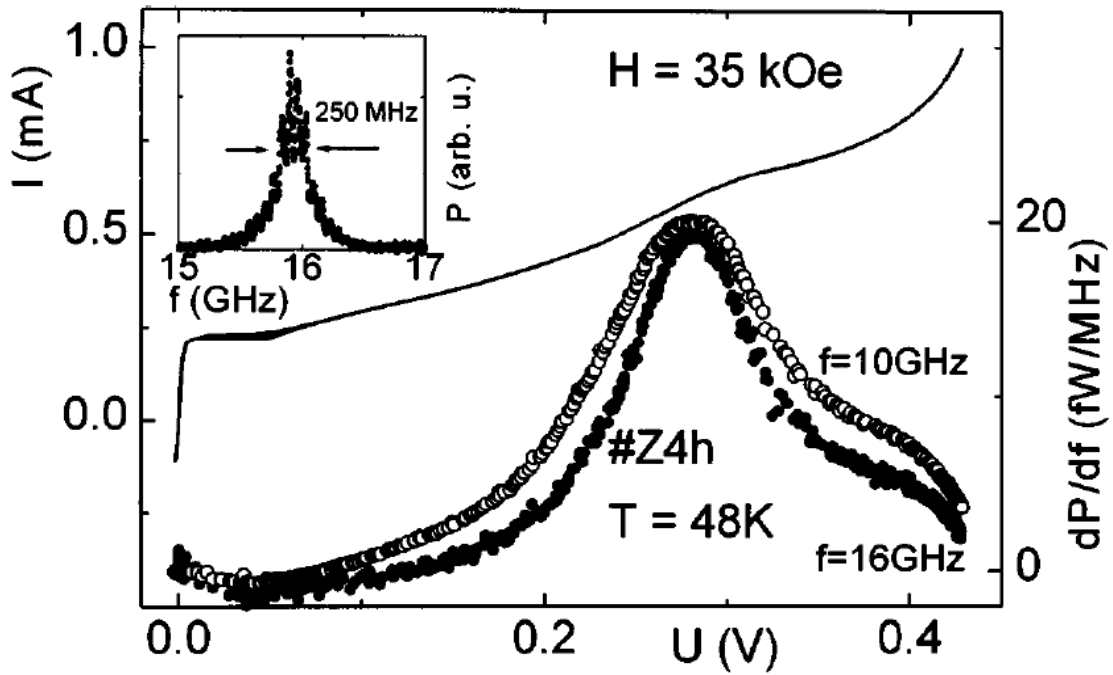

Figure 28. *IV* characteristics and non-Josephson microwave emission signals at 10 GHz (open circles) and 16 GHz (solid circles) for a magnetic field of 35 kOe. The curves were normalized to account for different coupling losses. Note that the form of the peak is almost the same for the two frequencies, showing that the non-Josephson signal is extremely broadband. The inset shows a typical spectrum of the Josephson signal, taken at 3 mV bias, which has a linewidth of only 250 MHz. [After G. Hechtfischer *et al*, Physical Review Letters 79, 1365-1368 (1997).].

Savel'ev *et al* (2005) calculated analytically the out-plane Cherenkov radiation in the spatially modulated layered high-$T_c$ superconductors [154]. They find that the electric field and magnetic field in the out-plane radiation has the same order of magnitude, thus the impedance mismatch problem encountered in the radiation from the *ac* and *bc*-plane is avoided.

*V.2 Radiation caused by quasiparticle injection*

Injection of quasiparticles into high-$T_c$ superconductors can also generate THz plasma oscillation and thus EM radiation. The basic idea is as follows [155]: the injected quasiparticles create a non-equilibrium distribution of electrons. These electrons condensate into Cooper pairs via recombination process, and thus excite Josephson plasma during the relaxation. The Josephson plasma then radiates EM wave into free space. The generation rate of Josephson plasma and the optimal condition for emission are analyzed by Shafranjuk and Tachiki (1999) [155].

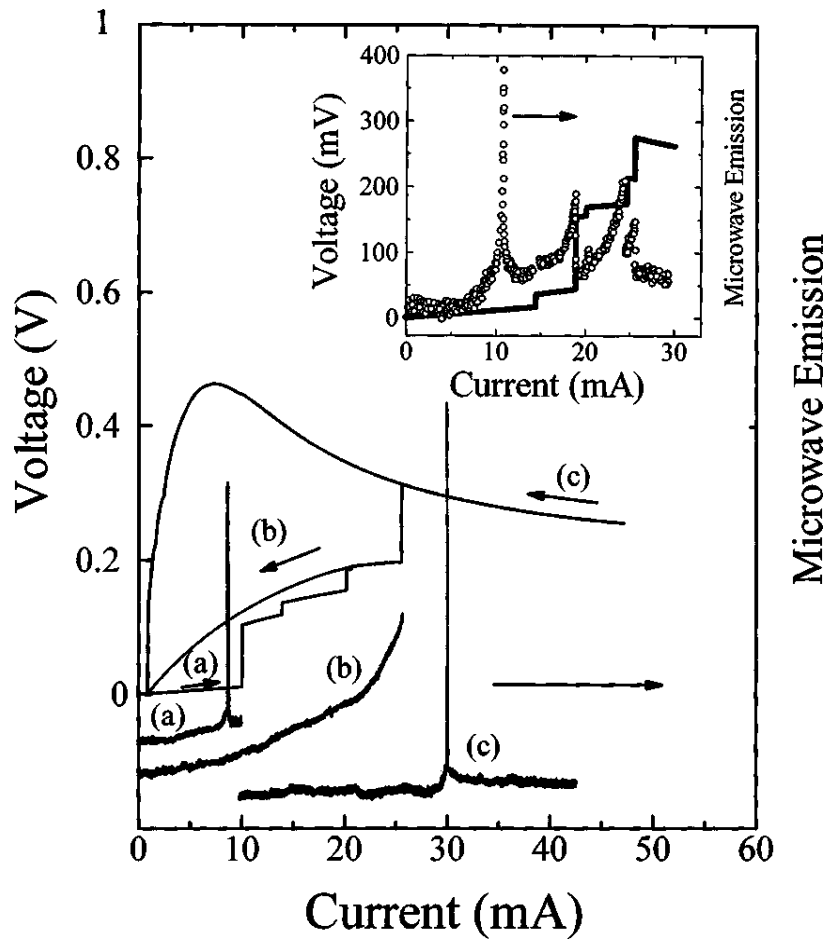

Figure 29. Microwave emissions at 11.6 GHz at 4.2 K together with the corresponding *IV* characteristics. For sections (a) and (c), the sharp peaks are observable, while only the broadband emission is found along curve (b). The inset shows a series of microwave emission spectra corresponding to the sweeping of several branches. [After K. Lee *et al*, Physical Review B 61, 3616-3619 (2000).].

In experiments [156-160], Iguchi *et al* injected quasiparticles into high-$T_c$ superconductors such as YBCO and BSCCO by a tunneling junction (Cu/I/S tunneling junction, S denotes superconductors and I denotes insulators). The radiation is detected by a superheterodyne mixer. The radiation power increases with the injection current and decreases with temperature, consistent with the theoretical calculations. Especially, in the experiment based on BSCCO[160], three different types of radiations are observed: the Josephson self-emission at low bias voltage with the voltage and frequency obeying the ac Josephson relation (curve a in Figure 29), the incoherent broadband emission due to injection of quasiparticles in the quasiparticle branch at high bias region (curve b in Figure 29), and the coherent emission in the negative-resistance region (curve c in Figure 29). This implies that the negative-resistance may play a role in the radiation as Gunn oscillators [161].

Kransov (2006, 2009) investigated the relaxation of non-equilibrium quasiparticles in cuprate superconductors both theoretically and experimentally [162, 163], and showed that the emission is amplified by cascade tunneling of quasiparticles similar to that in quantum cascade laser in semiconducting hetero structures [164-166], as depicted in Figure 30. The radiation frequency is estimated to cover the whole THz band and the efficiency is much higher than that of Josephson emission.

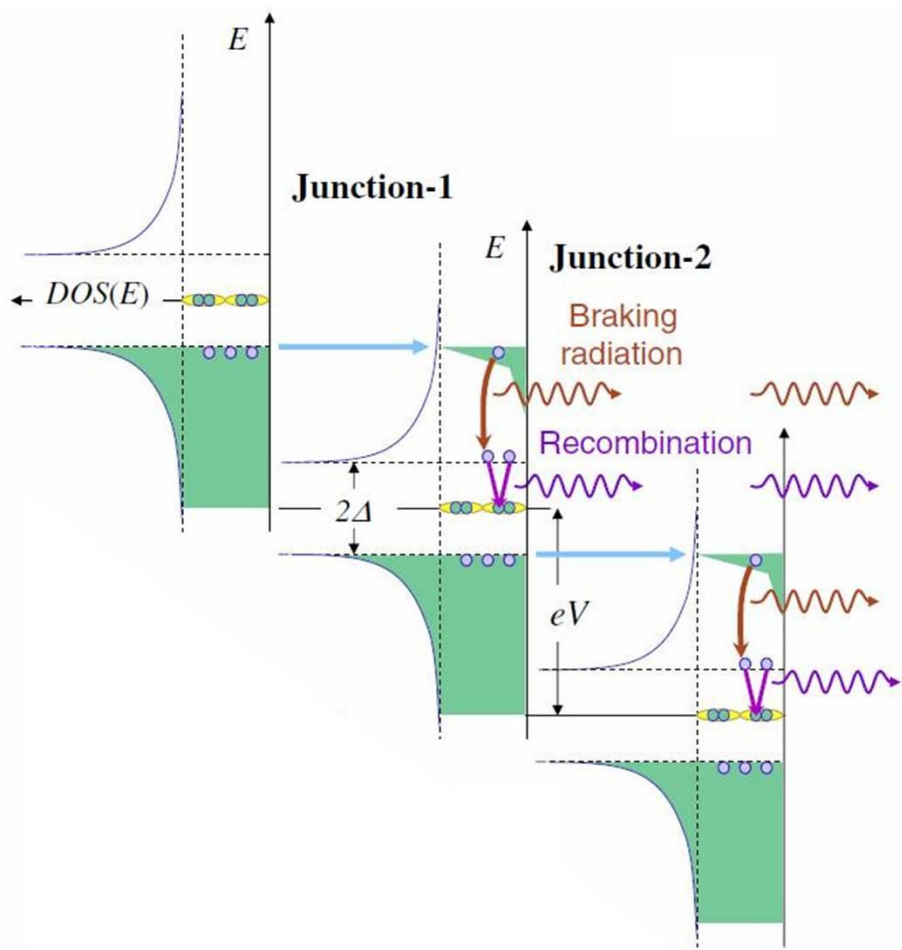

Figure 30. A schematic energy diagram of two stacked superconductor-insulator-superconductor (SIS) junctions biased at voltage $V$ per junction. For $V \geq 2\Delta/e$ tunneling results in non-equilibrium quasiparticle population in the empty band. Arrows indicate the most probable relaxation scenario of quasiparticles. The process is repeated in the second junction, resulting in cascade amplification of radiation, as indicated by wavy arrows. [After V. M. Krasnov, Physical Review Letters 97, 257003 (2006).]

*V.3 Radiation by moving Abrikosov vortex lattice*

The motion of Abrikosov vortex lattice can also generate EM radiation. It is observed that, imposing a rf current on the dc transport current, current steps appear in the *IV* curve in the flux flow regime [167-170]. It originates in the interaction between defects and the motion of vortices. The periodic passing of vortices through a barrier induced by defects is similar to a particle travelling in a washboard potential. Thus the current steps in flux flow regime have an origin similar to the Shapiro steps. By analogy with the radiation from Josephson junctions, inversely radiation from flux flow may be expected [171, 172].

The excitation of plasma by motion of Abrikosov vortices in isotropic superconductors was first analyzed by Koyama and Tachiki (1995) [173]. They find that the moving vortices with a constant velocity do not excite the plasma mode. Assuming a modulation of the velocity, they estimated the radiation power as high as $100\text{W/cm}^2$. A more realistic calculation was performed by Bulaevskii and Chudnovsky (2006) [171], where they considered the interaction between flux flow and disorder with appropriate boundary condition. The radiation frequency is $\omega_0 = 2\pi v / a$ up to a superconducting energy gap $\Delta/\hbar$, where $v$ is the velocity of vortices and $a$ the inter vortex separation. Assuming a rectangular configuration of vortices, the power is estimated to $0.1\text{W/cm}^2$ for $YBa_2Cu_3O_{7-\delta}$, and the efficiency is of order of $10^{-4}$. The flux flow in clean sample may also radiate when the vortices are annihilated at the edge, similar to the case when a soliton is moving in a Josephson junction.

**VI Conclusion**

In the present review, we have discussed the coherent terahertz radiation based on the intrinsic Josephson effects of the cuprate high-$T_c$ superconductors. No special attention has been paid to the $d$-wave symmetry of Cooper pairing in the tunneling process, which is supposed to be included in the temperature dependence of the quasiparticle conductance as a phenomenological parameter. It is noticed that the existence of quasiparticles due to the nodes in the superconductivity gap of $d$-wave symmetry makes the present technique available even at low temperature.

The setup of a mesa of single crystal under dc voltage bias in the absence of external magnetic field has been proved experimentally to be very useful. The radiation power increases rapidly from 1pW to $50\,\mu$W during the past several years. The achieved power is still much smaller than the theoretical prediction and further increase is expected in the near future. The radiation frequency ranges from 0.5THz to 2THz which already covers the terahertz band. Several problems still need to be addressed in order to make this device out of laboratory.

The present setup is double-edged. On one hand, the impedance mismatch renders the mesa as a cavity, and thus is helpful for synchronization and makes it possible to pump large energy into the system in the way of cavity resonance; on the other hand it reduces much the transmission coefficient. Since the strong radiation involves cavity resonance, the available frequency is fixed by the geometry. In order to achieve tunability of the radiation frequency, one needs to go beyond the present setup. Tailoring the mesa geometry and optimizing coupling between the oscillator and load deserves future investigation. Effort in this direction is also expected to results in enhancement of the radiation power.

The protocol of stimulating terahertz electromagnetic radiation in the presence of external magnetic field has a significant merit in frequency tunability. While analytical and numerical calculations reveal that the radiation from the current driven rectangle Josephson vortex lattice is quite strong, there is no clear experimental demonstration on the coherent radiation so far. It is probably due to that the rectangular vortex lattice is fragile, and distortions caused by thermodynamic fluctuations and quenched disorders easily suppress the plasma oscillation and diminish the coherent radiation.

As discussed in the present article, the intrinsic Josephson effects in high-$T_c$ superconductors can be used for generator of electromagnetic wave. They can also be useful for other purpose such as THz waveguide, filter and amplifier, and so on, which remain to be explored in the future.

**Acknowledgements**

It is the great pleasure of the authors to thank L. Bulaevskii, G. Crabtree, I. Iguchi, K. Kadowaki, A. Koshelev, T. Koyama, W. Kwok, L. Ozyuzer, N. Pederson, M. Tachiki, H.-B. Wang, T. Yamashida, U. Welp, and P.-H. Wu for insightful discussions. Calculations were performed on SR11000 (HITACHI) in NIMS. This work has been supported by World Premier Initiative (WPI) on Materials Nanoarchitectonics, Ministry of Education, Culture, Sports, Science and Technology (MEXT) of Japan, and by CREST Project, Japan Science and Technology Agency (JST).

coherence in the vortex state of $Bi_2Sr_2CaCu_2O_{8+\delta}$ probed by Josephson plasma resonance," Physical Review Letters **78**, 1972-1975 (1997).

92. A. E. Koshelev, L. N. Bulaevskii, and M. P. Maley, "Josephson plasma resonance as a structural probe of vortex liquid," Physical Review Letters **81**, 902-905 (1998).

93. G. Hechtfischer, R. Kleiner, A. V. Ustinov, and P. Müller, "Non-Josephson emission from intrinsic junctions in $Bi_2Sr_2CaCu_2O_{8+y}$: Cherenkov radiation by Josephson vortices," Physical Review Letters **79**, 1365-1368 (1997).

94. S. Savel'ev, A. L. Rakhmanov, and F. Nori, "Using Josephson vortex lattices to control terahertz radiation: Tunable transparency and terahertz photonic crystals," Physical Review Letters **94**, 157004 (2005).

95. S. Savel'ev, A. L. Rakhmanov, and F. Nori, "Josephson vortex lattices as scatterers of terahertz radiation: Giant magneto-optical effect and Doppler effect using terahertz tunable photonic crystals," Physical Review B **74**, 184512 (2006).

96. H. B. Wang, P. H. Wu, and T. Yamashita, "Stacks of intrinsic Josephson junctions singled out from inside $Bi_2Sr_2CaCu_2O_{8+x}$ single crystals," Appl Phys Lett **78**, 4010-4012 (2001).

97. V. M. Krasnov, A. Yurgens, D. Winkler, and P. Delsing, "Self-heating in small mesa structures," J Appl Phys **89**, 5578-5580 (2001).

98. J. C. Fenton, and C. E. Gough, "Heating in mesa structures," J Appl Phys **94**, 4665-4669 (2003).

99. V. N. Zavaritsky, "Essence of intrinsic tunneling: Distinguishing instrinsic features from artifacts," Physical Review B **72**, 094503 (2005).

100. V. M. Krasnov, M. Sandberg, and I. Zogaj, "In situ measurement of self-heating in intrinsic tunneling spectroscopy," Physical Review Letters **94**, 077003 (2005).

101. H. B. Wang, T. Hatano, T. Yamashita, P. H. Wu, and P. Müller, "Direct observation of self-heating in intrinsic Josephson junction array with a nanoelectrode in the middle," Appl Phys Lett **86**, 023504 (2005).

102. M. H. Bae, J. H. Choi, and H. J. Lee, "Heating-compensated constant-temperature tunneling measurements on stacks of $Bi_2Sr_2CaCu_2O_{8+x}$ intrinsic junctions," Appl Phys Lett **86**, 232502 (2005).

103. X. B. Zhu, Y. F. Wei, S. P. Zhao, G. H. Chen, H. F. Yang, A. Z. Jin, and C. Z. Gu, "Intrinsic tunneling spectroscopy of $Bi_2Sr_2CaCu_2O_{8+\delta}$: The junction-size dependence of self-heating," Physical Review B **73**, 224501 (2006).

104. K. E. Gray, A. E. Koshelev, C. Kurter, K. Kadowaki, T. Yamamoto, H. Minami, H. Yamaguchi, M. Tachiki, W. K. Kwok, and U. Welp, "Emission of Terahertz Waves From Stacks of Intrinsic Josephson Junctions," IEEE Transactions on Applied Superconductivity **19**, 886-890 (2009).